\documentclass[%
 reprint,
 amsmath,amssymb,
 aps,
prl,
superscriptaddress
]{revtex4-1}

 \pdfoutput=1

\usepackage{graphicx}
\usepackage{dcolumn}
\usepackage{bm}
\usepackage{verbatim}
\usepackage{xcolor}
\usepackage{cancel}

\usepackage{hyperref}

\hypersetup{
    bookmarks=false,         
    unicode=false,          
    pdftoolbar=false,        
    pdfmenubar=true,        
    pdffitwindow=false,     
    pdfstartview={FitH},    
    pdftitle={A Rigorous Theory of Prethermalization without Temperature},    
    pdfauthor={W. W. Ho, W. D. Roeck},     
    pdfsubject={},   
    pdfcreator={},   
    pdfproducer={}, 
    pdfkeywords={Prethermalization} {Driven systems}  , 
    pdfnewwindow=true,      
    colorlinks=true,       
    linkcolor=black,          
    citecolor=blue,        
    filecolor=magenta,      
    urlcolor=blue           
}

\renewcommand{\vec}[1]{\boldsymbol{\mathbf{#1}}}

\newcommand{\ww}[1]{\textcolor{red}{#1}}

\begin{document}

\title{
A Rigorous Theory of Prethermalization without Temperature
}
\author{Wen Wei Ho}
\affiliation{Department of Physics, Stanford University, Stanford, CA 94305, USA } 

\author{Wojciech De Roeck}
\affiliation{Instituut Theoretische Fysica, KU Leuven, 3001 Leuven, Belgium}

\begin{abstract}
    Prethermalization refers to the physical phenomenon where a system evolves toward some long-lived non-equilibrium steady state  before eventual thermalization sets in.
    One general scenario where this occurs is in driven systems with dynamics   governed by an effective Hamiltonian  (in some rotating frame), such that ergodicity of the latter is responsible for the approach to the prethermal state.
    %
    %
    This begs the question whether it is possible to have a   prethermal state not associated to any effective Hamiltonian. 
    Here, we answer this question in the affirmative. We exhibit a natural class of systems in which the prethermal state is defined by   emergent, global symmetries,  but where the dynamics that takes the system to this state has no additional conservation laws, in particular energy.  
    We explain how novel prethermal phases of matter can nevertheless emerge under such settings, distinct from those previously discussed.
\end{abstract}
\date{\today}
\maketitle

%
{\it Introduction.}---Recent years have seen dramatic progress towards understanding universal features of quantum many-body systems out of equilibrium. 
This has led to discoveries of exotic nonequilibrium physical phenomena, like ergodicity-breaking scenarios of many-body localization \cite{baskoMBL,gornyi2005interacting, husenandrev, RevModPhys.91.021001} and quantum many-body scars \cite{qmbs5,PeriodicOrbits}, as well as  novel   phases of matter   realized only in driven settings, like the discrete time crystal (DTC) \cite{PhysRevLett.116.250401, PhysRevLett.117.090402,yao2017discrete},   anomalous Floquet insulator \cite{PhysRevX.6.021013, PhysRevB.99.195133} and chiral Floquet phases \cite{PhysRevX.6.041070, PhysRevX.7.011018}. 


%
A key development in the theory of non-equilibrium many-body systems has been the establishment of the paradigm of prethermalization \cite{PhysRevLett.93.142002, kinoshita2006quantum, moeckel2008interaction, eckstein2009thermalization, gring2012relaxation, bertini2015prethermalization, PhysRevLett.116.120401, KUWAHARA201696, Abanin2017, PhysRevB.95.014112, howell2019asymptotic, mori2018floquet, rajak2018stability, PhysRevX.7.011018, PhysRevX.9.021027,reimann2019typicality,  huveneers2020prethermalization,rubio2020floquet}. 
This is the phenomenon where the system evolves towards a  long-lived, quasistationary state, which might exhibit interesting features not found in the true equilibrium state realized at extremely late times. Perhaps the cleanest incarnation of the phenomenon is exemplified by a  system  periodically driven at a high frequency: it has been   shown that despite time-translation symmetry being explicitly broken, there nevertheless exists a static, quasilocal energy operator $H_\text{eff}$, an `effective Hamiltonian', that is approximately conserved for times exponentially long in the driving frequency, and is moreover an effective generator of stroboscopic dynamics \cite{lazarides_das_14,dalessio_rigol_14,bukov2015universal,PhysRevLett.116.120401, KUWAHARA201696, Abanin2017, PhysRevB.95.014112}.
The system therefore   resists heating   towards the featureless, infinite-temperature state expected on grounds of entropy maximization in the absence of any global conservation laws, and instead equilibrates  to a thermal state  with well-defined temperature set by its (approximately conserved) energy, at least for such `prethermal times'.

The prethermal regime of a time-periodic (Floquet) system can in fact exhibit much richer structure than just conservation of energy. Ref.~\cite{PhysRevX.7.011026} identified a class of strong, high-frequency driving that leads to an additional   $\mathbb{Z}_n$ symmetry of   $H_\text{eff}$, whose presence underpins   the existence of Floquet prethermal phases. 
It is important to note that this symmetry is an emergent and robust one, and not tied to any exact, microscopic conservation laws nor a particular fine-tuned driving protocol.
 This result has recently been extended to systems driven with several mutually incommensurate frequencies -- so called quasiperiodically-driven systems --  such that {multiple} emergent $\mathbb{Z}_n$ symmetries can be engineered of the effective Hamiltonian \cite{dumitrescu2018logarithmically,PhysRevX.10.021032}. 
 Similar statements hold in time-independent settings where it has been shown how to robustly protect $U(1)$ conservation laws for long times, even in the presence of explicit symmetry-breaking terms \cite{Abanin2017, PhysRevX.10.021032}.
 %
%
 In all of the above cases, the properties of the prethermal state  are intrinsicially tied to the properties of an effective, static Hamiltonian description of dynamics (in some appropriate rotating frame). 
 

 In this Letter, we establish in a rigorous manner novel prethermalization scenarios where a strongly-driven quantum many-body system can exhibit  long-lived, emergent charge conservation, but without necessarily   energy conservation. 
 Concretely, we present results of (i)   long-lived $U(1)$-charge conservation in  Floquet systems, and (ii)   long-lived $\mathbb{Z}_n$-charge conservation in  quasiperiodically-driven systems, without reference to an effective, static Hamiltonian description of dynamics. 
 These are achieved in classes of systems containing some large energy scale $\nu$ well separated from all other energy scales in the system,   but which is not   borne out in the limit of high-frequency driving.
 %
 In particular, some or all of the drive frequencies   may be small, precluding an effective Hamiltonian construction like the Magnus expansion or its variants \cite{Blanes_2009, PhysRevLett.116.120401, KUWAHARA201696, Abanin2017, PhysRevB.95.014112, PhysRevX.10.021032}. 
 Multiple $U(1)$ or $\mathbb{Z}_n$ charge conservation may also be realized at the expense of increasing the number of fundamental frequencies of the drives. 
 In all cases, the prethermal timescale we   derive is superpolynomially long in $\nu$.

 Our results represent a rigorous realization of an exotic scenario of ``prethermalization without temperature''  coined by Ref.~\cite{PhysRevX.10.021046}, who envisaged a scenario where an  emergent charge conservation leads to 
  nontrivial dynamics for long times, despite dynamics occurring either (i) due to an effective Hamiltonian but at energies corresponding to high or infinite temperatures, or (ii)   in the absence of an effective Hamiltonian description such that temperature is not well defined in the prethermal state. Ref.~\cite{PhysRevX.10.021046}   analyzed the former situation.
Here, we   provide   general conditions showing that the latter scenario can in fact   occur. 
 Indeed, one of the nontrivial physical consequences of our work is that just the conservation of an emergent charge 
 in the prethermal state  is already sufficient to sharply define distinct prethermal  phases of matter, for example those distinguished by the absence or presence of a non-zero plateau of a local `order parameter'  in dynamics.
 %
This represents a novel class of prethermal phases different than has been previously discussed, for instance by \cite{PhysRevX.7.011026, PhysRevX.10.021032}, which are based on the existence of an effective Hamiltonian.

 {\it Key ideas.}---The general  setting  behind our theorems is encapsulated by the following class of driven many-body Hamiltonians  
 \begin{align}
     G(t) = \nu N + H(t),
     \label{eqn:G_setup}
 \end{align}
 where   $N$ is a term that has uniform spectral gaps which remain open in the thermodynamic limit 
 (for example, a Zeeman field  on quantum spins or a non-vanishing bandgap of lattice fermions, see \cite{PhysRevX.7.011018, Gulden_2020}).
 $H(t)$ represents  interactions or   couplings that depend periodically on time with frequency $\omega$. 
 Here, $\nu$ is the amplitude of $N$, taken to be much larger than all other local energy scales, which include  $\omega$ and the local bandwidth of $V(t)$ assumed to be bounded at any time by $J$. We do not need any relation of $\omega$ to $J$.  What we will show is that under an additional assumption of sufficient smoothness of the drive, there is a dressed version of $N$ that is conserved to exponentially long times in $\nu$. 

 \begin{figure}[t]
    \includegraphics[width = 0.5 \textwidth]{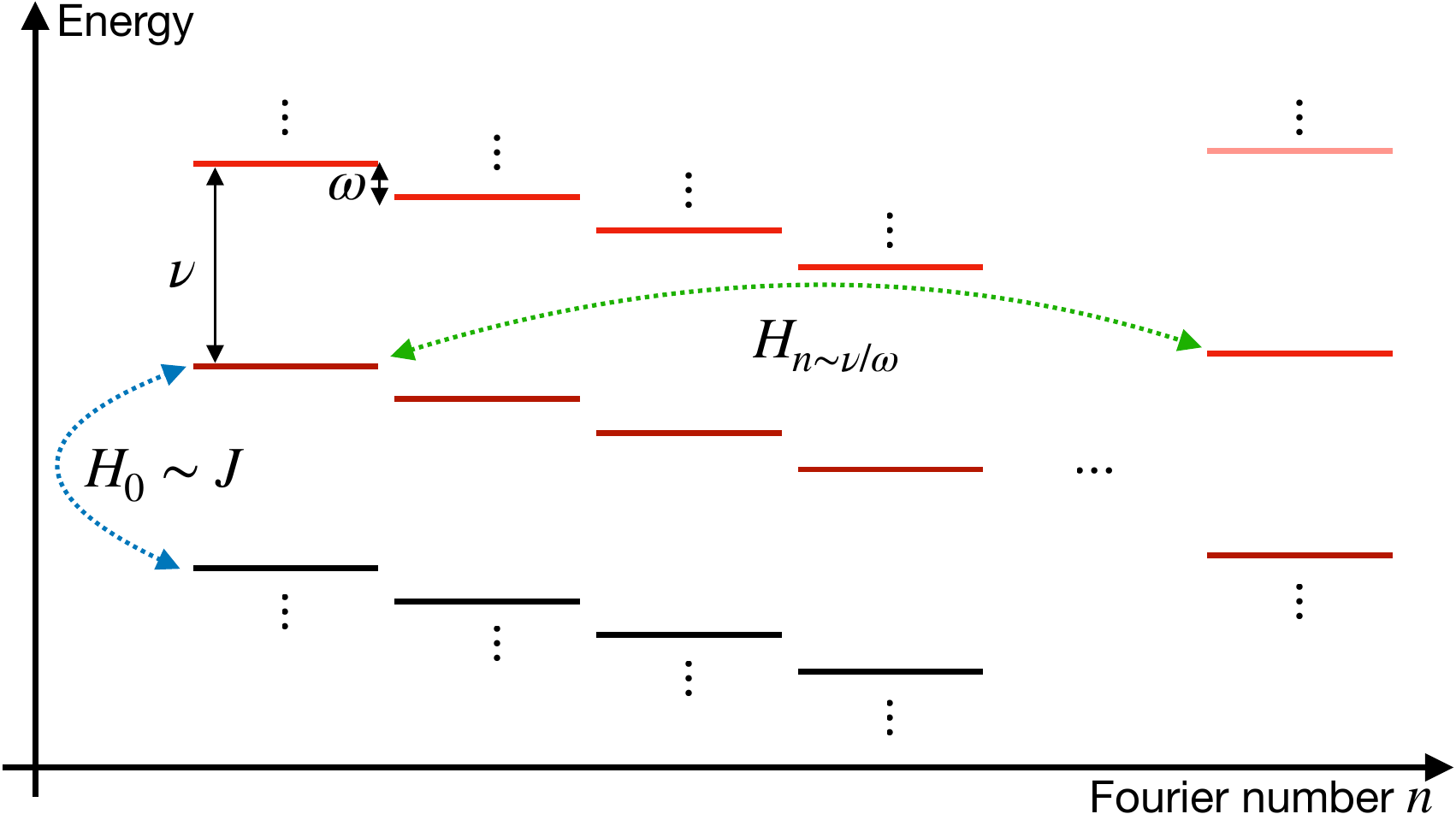}
\caption{Processes out of a given sector of $N$. The operator $\nu N$ has large spectral gaps  $\nu$, depicted as a tower of energies. There are processes effected by the Fourier mode $H_0$ (also present in the absence of a drive). The local nature of interactions entail they enter only with amplitude $J$, hence are off-resonant and heavily suppressed. In the presence of the drive there can be processes   involving involving multiples of the frequency $\omega$, in principle leading to a resonant coupling effected by $H_n$ with Fourier number $n$\,$\sim$\,$\nu/\omega$. Such processes can also be suppressed if the drive is smooth (local in Fourier space) so that the Fourier modes decay fast with $n$.
}
\label{fig:1}
\end{figure}

 To intuitively understand why this might be true, consider temporarily that $H(t)$ is time-independent. Then we are back to a previously considered situation \cite{Abanin2017} where we know there {\it is} an emergent charge which is approximately conserved. Schematically, the large spectral gaps of $\nu N$  entail that we can   `integrate' out processes  coupling different subspaces of $N$, leading to only effective couplings within them (this is akin to a Schrieffer-Wolff transformation \cite{Bravyi_2011}).
 As transitions are local in real space by assumption, such a procedure is always well-defined initially and can be carried out to high-orders $\sim$\,$\nu/J$ until a many-body resonance is encountered, which gives a bound on the rate of loss of conservation \cite{Abanin2017}. 

 Eq.~\eqref{eqn:G_setup} represents a generalization   in which there can be additional  processes that couple states separated in energy by multiples of the drive frequency $\omega$ (Fig.~\ref{fig:1}).  
In particular,  there can be direct, resonant transitions between  states belonging to different $N$ sectors  with energy difference $\nu$, with accompanying absorption or emission of $n$\,$\sim$\,$\nu/\omega$ quanta of drive energy $\omega$ effected by the Fourier modes $H_n$ of the   Hamiltonian. This can in principle result in a rapid loss of conservation of $N$. 
 %
 Our key observation  is that we can suppress such deleterious processes if we impose that $H(t)$ is sufficiently smooth in time, such that there is a   fast decay of $H_n$ with Fourier number $n$, i.e.~if we   impose locality in Fourier space. 
 For example, if the drive is analytic, then $H_n$ is at least exponentially small in $n$, leading to an exponential in $\nu/\omega$ suppression of the  direct transition amplitude. 
 Thus, we see how $N$ (or rather a dressed version) can once again be conserved for long times, controlled by the large factor $\nu/\max(J$\,$,$\,$\omega)$.
Our theorems  make concrete these considerations and extend them to more general charges as well as driving settings.

 {\it Statement of theorems.}---We    state here our formal theorems in a manner as self-contained as possible, relegating the complete mathematical details to the Supplemental Material (SM) \cite{SM}.  
 We consider a quantum many-body Hamiltonian $G$ on a lattice with locally bounded Hilbert space, for example of quantum spins or fermions. 
 We also take it to be parameterized by angles $\theta$\,$\in$\,$[0$\,$,$\,$2\pi)$ (Floquet case) or $\vec\theta$\,$=$\,$(\theta_1$\,$,$\,$\cdots$\,$,$\,$\theta_m)$\,$\in$\,$[0$\,$,$\,$2\pi)^m$ (quasiperiodic case, $m$\,$\geq$\,$2$). Indeed,  we can define, in a slight abuse of notation, a time-periodic Hamiltonian via $G(t)$\,$:=$\,$G(\theta_t)$  where $\theta_t$\,$=$\,$\omega t$\,$+$\,$\theta_0$\,$\mod$\,$2\pi$ ($\omega$: frequency, $\theta_0$: arbitrary initial phase), and a time-quasiperiodic one $G(t)$\,$:=$\,$G(\vec\theta_t)$ where $\vec\theta_t$\,$=$\,$\vec\omega$\,$t$\,$+$\,$\vec\theta_0$\,$\mod$\,$2\pi$ for each argument.  Here $\vec\omega$\,$=$\,$(\omega_1$\,$,$\,$\cdots$\,$,$\,$\omega_m)$ is a vector of frequencies and $\vec\theta_0$ an arbitrary vector of initial phases.  In both cases, our object of interest is the unitary propagator $U(t)$ satisfying the Schr\"{o}dinger equation $i\partial_tU(t)$\,$=$\,$G(t)U(t)$  with initial condition $U(0)$\,$=$\,$\mathbb{I}$.

 
 We furthermore assume the Hamiltonians $G(\theta)$\,$,$\,$G(\vec\theta)$ are sufficiently local in   real space and   smooth in $\theta,\vec\theta$ (this translates to the driven Hamiltonian $G(t)$ being  smooth in time). We measure this via   a local norm $\|G(\cdot)\|_\kappa$  parameterized by a decay constant $\kappa$\,$>$\,$0$, which takes into account the  decay of   local terms making up $G(\cdot)$ in both spatial extent and Fourier space (see SM for details \cite{SM}; similar norms were used in \cite{Abanin2017, PhysRevX.10.021032}). 
We can now state our first theorem. \\
 {\bf Theorem 1. Approximate $U(1)$-conservation in Floquet systems.} {\it Let $N$ (i) be a sum of local terms that mutually commute \& (ii) has integer spectrum, and $H(\theta)$ be a many-body Hamiltonian where $\theta$\,$\in$\,$[0,2\pi)$, with local norm $\|H(\theta)\|_{\kappa_0}$\,$<$\,$\infty$ for some $\kappa_0$\,$>$\,$0$.
 Let $\omega$\,$>$\,$0$ and define the local energy scale $\nu_0$\,$:=$\,$\max\{2\|H(\theta)\|_{\kappa_0},\omega\}$. We consider the Hamiltonian
 \begin{align}
     G(\theta) = \nu N + H(\theta)
     \label{eqn:G1}
 \end{align}
 (and correspondingly, dynamics under the time-periodic (Floquet) Hamiltonian $G(t)$\,$:=$\,$G(\theta_t)$ with fundamental frequency $\omega$), where the amplitude $\nu$ is assumed large, specifically $\nu$\,$>$\,$C\nu_0$ for some constant $C$ depending only on $\kappa_0$  but   not the volume of the system.
 Then, there is a small unitary $e^{A(\theta)}$ effected by a quasilocal, antihermitian operator $A(\theta)$, such that the   unitary propagator corresponding to $G(t)$   can be written
 \begin{align*}
     U(t) = e^{A(\theta_t)} \mathcal{T}\exp\left(-i \!\! \int_0^t \!\!\! ds \nu N + D(\theta_s) + V(\theta_s) \right)e^{-A(\theta_0)},
 \end{align*}
 where $\mathcal{T}$ represents time-ordering, and $D(\theta)$\,$,$\,$V(\theta)$ are quasilocal, many-body Hamiltonians satisfying
 \begin{align}
     & \|D(\theta) - \langle H (\theta) \rangle \|_{\kappa} \leq C'(\nu_0/\nu), \\
     & \| V(\theta)\|_{\kappa} \leq \nu_0 2^{-n_*}, \\
     & [D(\theta),N] = 0.
 \end{align}
 Here $\kappa$\,$=$\,$\kappa_0/4$, $\langle\cdot\rangle$ represents the symmetrization operation $\langle O(\theta)\rangle$\,$=$\,$\frac{1}{2\pi} \int_0^{2\pi} d\phi e^{i \phi N} O(\theta) e^{-i \phi N} $, and $n_*$\,$=$\,$\lfloor c (\nu/\nu_0)\rfloor$.  $C',c$ are   numerical constants independent of volume.
 }

Unpacking the theorem, it says that there is a small (close to identity) time-periodic change of frame such that  dynamics is  generated  by a time-periodic Hamiltonian $\nu N$\,$+$\,$D(\theta_t)$ which conserves $N$,  a $U(1)$ charge. Corrections to this (explicit symmetry-breaking terms $V(\theta_t)$) are very weak, being exponentially suppressed in $\nu$, and can be ignored.
This statement can be made precise for the case of local observables invoking Lieb-Robinson bounds \cite{Lieb1972}, see \cite{SM}.
One obvious consequence  is that in the laboratory frame, the dressed charge $\tilde{N}$\,$=$\,$e^{A(\theta_0)}N e^{-A(\theta_0)}$, which is a sum of quasilocal terms, is approximately conserved at stroboscopic times $t$\,$=$\,$\mathbb{Z}T$ ($T$: period) up to a prethermal time $\tau$ that is exponentially long in $\nu/\nu_0$ \cite{SM}. Relatedly, the original charge $N$ is approximately conserved for similar times, albeit up to a bounded error of $O(\nu_0/\nu)$.

Crucially, the theorem does not require any relation of the drive frequency $\omega$ to local energy scales of the system  $\|H(\theta)\|_{\kappa_0}$. In particular,   $\omega$ could  be smaller or even comparable to $\|H(\theta)\|_{\kappa_0}$, such that a further `high-frequency' (e.g.~Magnus)  expansion on the time-dependent $D(\theta_t)$ to obtain an effective Hamiltonian might not make sense. It is generally expected that there is then no notion of energy which is (approximately) conserved. Thus, our result pertains to one in which there is a long-lived emergent charge conservation in a driven system, without necessarily accompanying emergent energy conservation, as claimed.
 
Remark 1: while Theorem 1 specifies a set-up where $\nu$  is constant in time, we can actually apply it to a large class of cases where  $\nu$\,$=$\,$\nu(t)$ is time-periodic, see \cite{SM}. 
 Remark 2: We can extend our result to conservation of {\it multiple}  $U(1)$ charges, by upgrading $N$ to $r$ mutually commuting $U(1)$ charges $N_1,$\,$\cdots,$\,$N_r$, and promoting $\nu\mapsto\vec\nu$\,$=$\,$(\nu_1,$\,$\cdots,$\,$\nu_r)$. Similarly, we can promote $\theta \mapsto \vec\theta$ and achieve   results of $U(1)$ charge conservation in quasiperiodically-driven systems \cite{SM}.
 Remark 3: Technically speaking our theorem applies to systems  with interactions decaying at least exponentially with distance in real space, implicit   in the definition of local norm. This restriction may be   lifted to encompass long-range interactions combining techniques of \cite{PhysRevX.10.011043}.

 Let us provide  here a sketch of the proof; details are in \cite{SM}. 
 The proof technique relies on a rigorous implementation of Schrieffer-Wolff transformations, i.e.~many-body versions of Kolmogorov-Arnold-Moser (KAM) or Nekoroshev techniques, and it goes back at least to \cite{datta1996low}; we have been especially influenced by 
 \cite{benettin1988nekhoroshev,cuneo2017energy,giorgilli2015extensive, frohlich1986localization, imbrie2016many,Abanin2017,  PhysRevX.10.021032}. 
 The logic is to iteratively renormalize the Hamiltonian 
 such that terms  off-diagonal in $N$ have reduced amplitude, possible because  the large energy scale $\nu$ allows   to `integrate out' such processes. 
 More precisely, we introduce  a sequence of small unitaries $e^{-A_0(\theta)}$\,$,$\,$e^{-A_1(\theta)},\cdots$ with antihermitian $A_0(\theta)$\,$,$\,$A_1(\theta)$\,$,$\,$\cdots$, so that at step $n+1$ we have a rotated Hamiltonian
 \begin{align*}
    \nu N + H_{n+1}(\theta) \equiv e^{-A_n(\theta)}(& \nu N + H_n(\theta)  - i \omega  \partial_\theta) e^{A_n(\theta)}.
\end{align*}
(The original Hamiltonian $H(\theta)$ is labeled $H_0(\theta)$). 
If the Hamiltonian were time-independent, the last term $i\omega$\,$e^{-A_0(\theta)}$\,$\partial_{\theta}$\,$e^{A_0(\theta)}$, a `gauge potential', would not exist and we would reduce to the analysis of \cite{Abanin2017}. 
There it was shown how a choice of $A_n$ satisfying $[\nu N, A_n]$\,$=$\,$-V_n$ performed the renormalization, where $V_n$\,$:=$\,$H_n$\,$-$\,$\langle H_n \rangle$ is the off-diagonal part of $H_n$ (we term the diagonal part  $D_n$\,$:=$\,$\langle H_n \rangle$). 
Indeed, $A_n$ is $1/\nu$ small, so expanding $\nu N$\,$+$\,$H_{n+1}$  (still pretending it is time-independent) yields $\nu N$\,$+$\,$D_n$\,$+$\,$\cancel{V_n-[A_n,\nu N]}$\,$+$\,$O(\nu^{-1})$ and we   see the strength of $V_{n+1}$ in $H_{n+1}$ is reduced by  a factor $1/\nu$ relative to $V_n$. 
Of course,  all objects are many-body operators and so we should  measure amplitudes via the local norm $\|\cdot\|_{\kappa}$. The price to pay of the renormalization  is a slight decrease of the locality of the   Hamiltonian $H_{n+1} $  resulting in a smaller reduction factor than just $1/\nu$.

In our present case, we must   account for the effect of the gauge potential.
%
Now,  suppose we   continue to choose the previous solution of $A_n$  for each  $\theta$, simply promoting $A_n$\,$\mapsto$\,$A_n(\theta)$.
Then the key point is this: should $V_n(\theta)$ be a smooth function of $\theta$, so will $A_n(\theta)$.  Its derivative will then be bounded, and we can estimate
$\|\omega$\,$e^{-A_n(\theta)}$\,$\partial_{\theta}$\,$e^{A_n(\theta)}$\,$\|_\text{local}$\,$\sim$\,$\|\partial_\theta A_n(\theta)\|_\text{local}$\,$\omega$\,$\lesssim  \text{const.}$\,$\times$\,$\|V_n(\theta)\|_\text{local}$\,$\omega$\,$/$\,$\nu$ which will be  small should $\omega$\,$\ll$\,$\nu$. That is to say, the size of the off-diagonal terms  $V_{n+1}(\theta)$ will still be $\sim$\,$1/\nu$ smaller than $V_n(\theta)$ even in the time-dependent scenario. 
Iterating the procedure up to the optimal order $n_*$ then yields Theorem 1.

A slight modification leads us to:\\
 {\bf Theorem 2. Approximate $\mathbb{Z}_n$-charge conservation in quasiperiodically-driven systems away from the high-frequency limit.}
{\it
Consider $N$ (i) a sum of local terms that  mutually commute and (ii) has integer eigenvalue spacings. Fix a non-zero integer $n$ and let $H(\vec\theta)$ be a many-body Hamiltonian on $\vec\theta$\,$\in$\,$[0$\,$,$\,$2\pi)^2$, assuming that $\|H(\vec\theta)\|_{\kappa_0}$ for some $\kappa_0$\,$>$\,$0$. 
We introduce a frequency vector $\vec\omega$\,$=$\,$(\nu$\,$,$\,$\omega)$, define $\nu_0$\,$:=$\,$\max\{2\|H(\vec\theta)\|_{\kappa_0},\omega\}$, and consider the Hamiltonian
\begin{align}
    G(\vec\theta) = \frac{\nu}{n} N + H(\vec\theta)
    \label{eqn:G2}
\end{align}
(and corresponding, the time-quasiperiodic Hamiltonian $G(t)$\,$:=$\,$G(\vec\theta_t)$). We take $\nu$\,$>$\,$C\nu_0$ for some constant $C$ depending on $\kappa_0$ but not on the system's volume.
Then, there is a small time-quasiperiodic unitary $e^{A(\vec\theta_t)}$ effected by a quasilocal, antihermitian operator $A(\vec\theta)$   such that the unitary propagator can be written  
\begin{align*}
    U(t) = e^{A(\vec\theta_t)} \mathcal{T} \exp\left(-i\!\! \int_0^t \!\!\! ds \frac{\nu}{n}N + D(\vec\theta_t) + V(\vec\theta_t) \right) e^{-A(\vec\theta_0)},
\end{align*}
where $D(\vec\theta), V(\vec\theta)$ are quasilocal Hamiltonians satisfying
\begin{align}
    & \| D(\vec\theta) - \langle H(\vec\theta)\rangle \|_{\kappa} \leq C'(\nu_0/\nu), \\
    & \| V(\vec\theta) \|_{\kappa} \leq \nu_0 2^{-n_*}, \\
    & [D(\vec\theta),g] = 0.
\end{align}
Here   $D(\vec\theta)$\,$=$\,$D'(\theta_2)$ has   dependence only on $\theta_2$\,$\in$\,$[0,2\pi)$, 
  $\kappa$\,$=$\,$\kappa_0/4$,  $g$\,$=$\,$e^{i \frac{2\pi}{n}N }$ is a generator of the $\mathbb{Z}_n$ group satisfying $g^n$\,$=$\,$\mathbb{I}$,   $\langle \cdot \rangle$ is the symmetrization operation
  $\langle O(\vec\theta)\rangle$\,$=$\,$\frac{1}{2\pi n}\int_0^{2\pi n} d\theta_1 e^{-i \frac{\theta_1}{n}N}O(\vec\theta)e^{i \frac{\theta_1}{n}N}$, 
  $n_*$\,$=$\,$\lfloor c(\nu/\nu_0)\rfloor$, and   $C',c$ are numerical constants.
}

 Theorem 2 (proof given in the SM \cite{SM})) specifies that in this strongly-driven set-up, 
 there is a small time-quasiperiodic change of frame where dynamics is generated by a $\mathbb{Z}_n$-symmetric, but now {\it time-periodic} (in $T_2$\,$=$\,$2\pi/\omega$) Hamiltonian $D'(\omega t$\,$+$\,$(\theta_0)_2)$, with small corrections. 
 Therefore, similar to Theorem 1, the Heisenberg time evolution of a local operator  is essentially  governed just by this symmetric time-dependent Hamiltonian, for exponentially long times \cite{SM}. 
Note that in this scenario, we have utilized that one of the drive frequencies, $\nu$, is the large energy scale, while the other drive frequency $\omega$ need not be: it can again be comparable to or smaller than local energy scales. This is thus not captured by the   `high-frequency' driving regime of \cite{PhysRevX.10.021032} and hence a different kind of $\mathbb{Z}_n$ symmetry conservation from the one identified there. 
We remark that  the amplitude of $N$ in Eq.~\eqref{eqn:G2} can also be made time-periodic in $T_1$\,$=$\,$2\pi/\nu$ as long as its time-average equals $\nu/n$. Moreover, Theorem 2 can be upgraded to encompass multiple long-lived  emergent $\mathbb{Z}_n$ charges  \cite{SM}.  

{\it Discussion.}---We now spell out the physical consequences, focusing on the case of an emergent $U(1)$-charge conservation in a Floquet system without energy conservation. 
The prethermal state reached (after a relaxation time  $t_r$\,$\sim$\,$\nu_0^{-1}$, which is much shorter than the prethermal time $\tau$\,$\sim$\,$\nu_0^{-1}$\,$e^{c(\nu/\nu_0)}$) is simply one determined by a fixed density of the conserved charge $N$.
 Despite this apparent simplicity, this setting {\it does} allow for distinct dynamical phases with a dynamical phase transition separating them.
 %
 %
Indeed, consider as a paradigmatic case
\begin{align}
N= -\frac{1}{2}\sum_{\langle i j \rangle } \sigma_i^z\sigma_j^z,
\end{align}
where $\sigma^z$ are spin-1/2 Pauli-matrices and the summation is over nearest-neighbors on a  lattice, i.e.~the classical Ising Hamiltonian, which can be understood as  the total number of domain walls in the system. 
In an equilibrium ensemble determined by domain wall density $n$\,$\approx$\,$N/V$  with $V$ the volume, we have ferromagnetic ordering if $n<$\,$n_c(d)$ a critical   density, provided spatial dimension $d$\,$>$\,$1$. This ordering stems from the   $\mathbb{Z}_2$ (spin-flip) symmetry that $N$ additionally possesses. 

Now, the key point is that such ordering will   still be preserved dynamically, even if the time-dependent Hamiltonian $D(\theta_t)$ driving dynamics is {\it not} $\mathbb{Z}_2$-symmetric, as long as it {\it does} preserve $N$. 
%
%
%
%
In particular, we could have $D(\theta)$\,$=$\,$h(\theta)\sum_i\sigma^z_i$.
This stability is due to the kinetic barrier that separates the positive and negative magnetization (as measured by $N_m$\,$=$\,$\sum_i$\,$\sigma^z_i$) sectors within the ensemble of constant $N$. Indeed, 
one can argue that if the   domain wall density is low enough, it is difficult in $d$\,$>$\,$1$ to change the total magnetization of the system significantly without also changing the number of domain walls  at the same time. 
As our Theorem guarantees $N$ is  (approximately) conserved in dynamics, it means that an initial state with low enough domain wall density and large net magnetization will have its magnetization survive up to at least the parametrically-long prethermal timescale, while a state with high domain wall density and large net magnetization will have its magnetization decay rapidly. Furthermore, such   behavior is robust to changes in the drive protocol, justifying their terminology as realizing prethermal `phases of matter'.
%

Our Theorem can also be directly applied to constrain dynamics in   experimentally-relevant systems. Consider  an ensemble  of Rydberg atoms interacting via strong, repulsive Van der Waals forces between Rydberg states \cite{Labuhn_2016, Lukin1, PhysRevLett.120.180502, Browaeys_2020}. The effective Hamiltonian is one acting on a collection of two-level systems spanned by states $|g\rangle$\,$,$\,$|r\rangle$:
\begin{align}
    H = \frac{\Omega(t)}{2} \sum_i \sigma^x_i - \Delta(t) \sum_i n_i + \sum_{i<j} V_{ij} n_i n_j,
\end{align}
where $\sigma^x_i$\,$=$\,$|g\rangle_i\langle r|$\,$+$\,$\text{h.c.}$,  $n_i$\,$=$\,$|r \rangle_i\langle r|$, and $V_{ij}$\,$\propto$\,$\frac{1}{|i-j|^6}$. The atoms can be    arranged in such a way that the interaction between nearest-neighbor pairs $\langle ij\rangle$ is dominant, so we can identify $\nu N$\,$=$\,$\sum_{\langle ij \rangle}V_{ij}n_in_j$.
Now it is natural to argue that due to the large separation of energy scales we can effectively work with states with definite $N$; for $N$\,$=$\,$0$ this is the so-called Rydberg-blockaded regime (neighboring atoms cannot be simultaneously excited). 
In the case when the Rabi-frequency $\Omega(t)$ and detuning $\Delta(t)$ are both time-independent, the rigorous justification behind this (as well timescales of the description) is covered by   Theorems in \cite{Abanin2017, PhysRevX.10.011043}. Our present Theorems guarantee that this intuition in fact continues to hold for a large class of time-dependent scenarios, thereby allowing for an analysis of dynamics still within the Rydberg-blockaded space.

Lastly, let us briefly comment on extensions beyond our work.  Besides having a large energy scale, a key idea was that the drive should be smooth. 
 Such a treatment hence excludes step-drives, for example one where $N$'s amplitude varies as $+2/3\nu$ for half a period and $-1/3\nu$ for the other half, but is otherwise  the dominant energy scale instantaneously. 
 The work of \cite{deroeck2019slow} covers such a scenario  and presents similar results as us (albeit with different assumptions), indicating a prethermalization mechanism   different from the one identified here. 


{\it Conclusion.}---We have exhibited the occurrence of a novel prethermalization scenario. Its distinguishing feature is that the prethermal state is defined by a conserved, emergent charge, a $U(1)$ or $\mathbb{Z}_n$-charge, without any reference to an effective, static Hamiltonian. Despite its simplicity, this setting already allows for nontrivial examples of dynamical phases.

\begin{acknowledgments}
{\it Acknowledgments.---} W.W.H.~is supported in part by the Stanford Institute for Theoretical Physics at Stanford University. W.D.R. acknowledges the support of the Flemish Research Fund FWO undergrants
G098919N and G076216N, and the support of KULeuven
University under internal grant C14/16/062.
\end{acknowledgments}

\bibliography{refs}

\begin{thebibliography}{54}%
\makeatletter
\providecommand \@ifxundefined [1]{%
 \@ifx{#1\undefined}
}%
\providecommand \@ifnum [1]{%
 \ifnum #1\expandafter \@firstoftwo
 \else \expandafter \@secondoftwo
 \fi
}%
\providecommand \@ifx [1]{%
 \ifx #1\expandafter \@firstoftwo
 \else \expandafter \@secondoftwo
 \fi
}%
\providecommand \natexlab [1]{#1}%
\providecommand \enquote  [1]{``#1''}%
\providecommand \bibnamefont  [1]{#1}%
\providecommand \bibfnamefont [1]{#1}%
\providecommand \citenamefont [1]{#1}%
\providecommand \href@noop [0]{\@secondoftwo}%
\providecommand \href [0]{\begingroup \@sanitize@url \@href}%
\providecommand \@href[1]{\@@startlink{#1}\@@href}%
\providecommand \@@href[1]{\endgroup#1\@@endlink}%
\providecommand \@sanitize@url [0]{\catcode `\\12\catcode `\$12\catcode
  `\&12\catcode `\#12\catcode `\^12\catcode `\_12\catcode `\%12\relax}%
\providecommand \@@startlink[1]{}%
\providecommand \@@endlink[0]{}%
\providecommand \url  [0]{\begingroup\@sanitize@url \@url }%
\providecommand \@url [1]{\endgroup\@href {#1}{\urlprefix }}%
\providecommand \urlprefix  [0]{URL }%
\providecommand \Eprint [0]{\href }%
\providecommand \doibase [0]{http://dx.doi.org/}%
\providecommand \selectlanguage [0]{\@gobble}%
\providecommand \bibinfo  [0]{\@secondoftwo}%
\providecommand \bibfield  [0]{\@secondoftwo}%
\providecommand \translation [1]{[#1]}%
\providecommand \BibitemOpen [0]{}%
\providecommand \bibitemStop [0]{}%
\providecommand \bibitemNoStop [0]{.\EOS\space}%
\providecommand \EOS [0]{\spacefactor3000\relax}%
\providecommand \BibitemShut  [1]{\csname bibitem#1\endcsname}%
\let\auto@bib@innerbib\@empty
\bibitem [{\citenamefont {Basko}\ \emph {et~al.}(2006)\citenamefont {Basko},
  \citenamefont {Aleiner},\ and\ \citenamefont {Altshuler}}]{baskoMBL}%
  \BibitemOpen
  \bibfield  {author} {\bibinfo {author} {\bibfnamefont {D.}~\bibnamefont
  {Basko}}, \bibinfo {author} {\bibfnamefont {I.}~\bibnamefont {Aleiner}}, \
  and\ \bibinfo {author} {\bibfnamefont {B.}~\bibnamefont {Altshuler}},\ }\href
  {\doibase https://doi.org/10.1016/j.aop.2005.11.014} {\bibfield  {journal}
  {\bibinfo  {journal} {Annals of Physics}\ }\textbf {\bibinfo {volume}
  {321}},\ \bibinfo {pages} {1126 } (\bibinfo {year} {2006})}\BibitemShut
  {NoStop}%
\bibitem [{\citenamefont {Gornyi}\ \emph {et~al.}(2005)\citenamefont {Gornyi},
  \citenamefont {Mirlin},\ and\ \citenamefont
  {Polyakov}}]{gornyi2005interacting}%
  \BibitemOpen
  \bibfield  {author} {\bibinfo {author} {\bibfnamefont {I.~V.}\ \bibnamefont
  {Gornyi}}, \bibinfo {author} {\bibfnamefont {A.~D.}\ \bibnamefont {Mirlin}},
  \ and\ \bibinfo {author} {\bibfnamefont {D.~G.}\ \bibnamefont {Polyakov}},\
  }\href {\doibase 10.1103/PhysRevLett.95.206603} {\bibfield  {journal}
  {\bibinfo  {journal} {Phys. Rev. Lett.}\ }\textbf {\bibinfo {volume} {95}},\
  \bibinfo {pages} {206603} (\bibinfo {year} {2005})}\BibitemShut {NoStop}%
\bibitem [{\citenamefont {Nandkishore}\ and\ \citenamefont
  {Huse}(2015)}]{husenandrev}%
  \BibitemOpen
  \bibfield  {author} {\bibinfo {author} {\bibfnamefont {R.}~\bibnamefont
  {Nandkishore}}\ and\ \bibinfo {author} {\bibfnamefont {D.~A.}\ \bibnamefont
  {Huse}},\ }\href {\doibase 10.1146/annurev-conmatphys-031214-014726}
  {\bibfield  {journal} {\bibinfo  {journal} {Annual Review of Condensed Matter
  Physics}\ }\textbf {\bibinfo {volume} {6}},\ \bibinfo {pages} {15} (\bibinfo
  {year} {2015})}\BibitemShut {NoStop}%
\bibitem [{\citenamefont {Abanin}\ \emph {et~al.}(2019)\citenamefont {Abanin},
  \citenamefont {Altman}, \citenamefont {Bloch},\ and\ \citenamefont
  {Serbyn}}]{RevModPhys.91.021001}%
  \BibitemOpen
  \bibfield  {author} {\bibinfo {author} {\bibfnamefont {D.~A.}\ \bibnamefont
  {Abanin}}, \bibinfo {author} {\bibfnamefont {E.}~\bibnamefont {Altman}},
  \bibinfo {author} {\bibfnamefont {I.}~\bibnamefont {Bloch}}, \ and\ \bibinfo
  {author} {\bibfnamefont {M.}~\bibnamefont {Serbyn}},\ }\href {\doibase
  10.1103/RevModPhys.91.021001} {\bibfield  {journal} {\bibinfo  {journal}
  {Rev. Mod. Phys.}\ }\textbf {\bibinfo {volume} {91}},\ \bibinfo {pages}
  {021001} (\bibinfo {year} {2019})}\BibitemShut {NoStop}%
\bibitem [{\citenamefont {Turner}\ \emph {et~al.}(2018)\citenamefont {Turner},
  \citenamefont {Michailidis}, \citenamefont {Abanin}, \citenamefont {Serbyn},\
  and\ \citenamefont {Papi\ifmmode~\acute{c}\else \'{c}\fi{}}}]{qmbs5}%
  \BibitemOpen
  \bibfield  {author} {\bibinfo {author} {\bibfnamefont {C.~J.}\ \bibnamefont
  {Turner}}, \bibinfo {author} {\bibfnamefont {A.~A.}\ \bibnamefont
  {Michailidis}}, \bibinfo {author} {\bibfnamefont {D.~A.}\ \bibnamefont
  {Abanin}}, \bibinfo {author} {\bibfnamefont {M.}~\bibnamefont {Serbyn}}, \
  and\ \bibinfo {author} {\bibfnamefont {Z.}~\bibnamefont
  {Papi\ifmmode~\acute{c}\else \'{c}\fi{}}},\ }\href {\doibase
  10.1038/s41567-018-0137-5} {\bibfield  {journal} {\bibinfo  {journal} {Nature
  Physics}\ }\textbf {\bibinfo {volume} {14}},\ \bibinfo {pages} {745}
  (\bibinfo {year} {2018})}\BibitemShut {NoStop}%
\bibitem [{\citenamefont {Ho}\ \emph {et~al.}(2019)\citenamefont {Ho},
  \citenamefont {Choi}, \citenamefont {Pichler},\ and\ \citenamefont
  {Lukin}}]{PeriodicOrbits}%
  \BibitemOpen
  \bibfield  {author} {\bibinfo {author} {\bibfnamefont {W.~W.}\ \bibnamefont
  {Ho}}, \bibinfo {author} {\bibfnamefont {S.}~\bibnamefont {Choi}}, \bibinfo
  {author} {\bibfnamefont {H.}~\bibnamefont {Pichler}}, \ and\ \bibinfo
  {author} {\bibfnamefont {M.~D.}\ \bibnamefont {Lukin}},\ }\href {\doibase
  10.1103/PhysRevLett.122.040603} {\bibfield  {journal} {\bibinfo  {journal}
  {Phys. Rev. Lett.}\ }\textbf {\bibinfo {volume} {122}},\ \bibinfo {pages}
  {040603} (\bibinfo {year} {2019})}\BibitemShut {NoStop}%
\bibitem [{\citenamefont {Khemani}\ \emph {et~al.}(2016)\citenamefont
  {Khemani}, \citenamefont {Lazarides}, \citenamefont {Moessner},\ and\
  \citenamefont {Sondhi}}]{PhysRevLett.116.250401}%
  \BibitemOpen
  \bibfield  {author} {\bibinfo {author} {\bibfnamefont {V.}~\bibnamefont
  {Khemani}}, \bibinfo {author} {\bibfnamefont {A.}~\bibnamefont {Lazarides}},
  \bibinfo {author} {\bibfnamefont {R.}~\bibnamefont {Moessner}}, \ and\
  \bibinfo {author} {\bibfnamefont {S.~L.}\ \bibnamefont {Sondhi}},\ }\href
  {\doibase 10.1103/PhysRevLett.116.250401} {\bibfield  {journal} {\bibinfo
  {journal} {Phys. Rev. Lett.}\ }\textbf {\bibinfo {volume} {116}},\ \bibinfo
  {pages} {250401} (\bibinfo {year} {2016})}\BibitemShut {NoStop}%
\bibitem [{\citenamefont {Else}\ \emph {et~al.}(2016)\citenamefont {Else},
  \citenamefont {Bauer},\ and\ \citenamefont {Nayak}}]{PhysRevLett.117.090402}%
  \BibitemOpen
  \bibfield  {author} {\bibinfo {author} {\bibfnamefont {D.~V.}\ \bibnamefont
  {Else}}, \bibinfo {author} {\bibfnamefont {B.}~\bibnamefont {Bauer}}, \ and\
  \bibinfo {author} {\bibfnamefont {C.}~\bibnamefont {Nayak}},\ }\href
  {\doibase 10.1103/PhysRevLett.117.090402} {\bibfield  {journal} {\bibinfo
  {journal} {Phys. Rev. Lett.}\ }\textbf {\bibinfo {volume} {117}},\ \bibinfo
  {pages} {090402} (\bibinfo {year} {2016})}\BibitemShut {NoStop}%
\bibitem [{\citenamefont {Yao}\ \emph {et~al.}(2017)\citenamefont {Yao},
  \citenamefont {Potter}, \citenamefont {Potirniche},\ and\ \citenamefont
  {Vishwanath}}]{yao2017discrete}%
  \BibitemOpen
  \bibfield  {author} {\bibinfo {author} {\bibfnamefont {N.~Y.}\ \bibnamefont
  {Yao}}, \bibinfo {author} {\bibfnamefont {A.~C.}\ \bibnamefont {Potter}},
  \bibinfo {author} {\bibfnamefont {I.-D.}\ \bibnamefont {Potirniche}}, \ and\
  \bibinfo {author} {\bibfnamefont {A.}~\bibnamefont {Vishwanath}},\ }\href
  {\doibase 10.1103/PhysRevLett.118.030401} {\bibfield  {journal} {\bibinfo
  {journal} {Phys. Rev. Lett.}\ }\textbf {\bibinfo {volume} {118}},\ \bibinfo
  {pages} {030401} (\bibinfo {year} {2017})}\BibitemShut {NoStop}%
\bibitem [{\citenamefont {Titum}\ \emph {et~al.}(2016)\citenamefont {Titum},
  \citenamefont {Berg}, \citenamefont {Rudner}, \citenamefont {Refael},\ and\
  \citenamefont {Lindner}}]{PhysRevX.6.021013}%
  \BibitemOpen
  \bibfield  {author} {\bibinfo {author} {\bibfnamefont {P.}~\bibnamefont
  {Titum}}, \bibinfo {author} {\bibfnamefont {E.}~\bibnamefont {Berg}},
  \bibinfo {author} {\bibfnamefont {M.~S.}\ \bibnamefont {Rudner}}, \bibinfo
  {author} {\bibfnamefont {G.}~\bibnamefont {Refael}}, \ and\ \bibinfo {author}
  {\bibfnamefont {N.~H.}\ \bibnamefont {Lindner}},\ }\href {\doibase
  10.1103/PhysRevX.6.021013} {\bibfield  {journal} {\bibinfo  {journal} {Phys.
  Rev. X}\ }\textbf {\bibinfo {volume} {6}},\ \bibinfo {pages} {021013}
  (\bibinfo {year} {2016})}\BibitemShut {NoStop}%
\bibitem [{\citenamefont {Nathan}\ \emph {et~al.}(2019)\citenamefont {Nathan},
  \citenamefont {Abanin}, \citenamefont {Berg}, \citenamefont {Lindner},\ and\
  \citenamefont {Rudner}}]{PhysRevB.99.195133}%
  \BibitemOpen
  \bibfield  {author} {\bibinfo {author} {\bibfnamefont {F.}~\bibnamefont
  {Nathan}}, \bibinfo {author} {\bibfnamefont {D.}~\bibnamefont {Abanin}},
  \bibinfo {author} {\bibfnamefont {E.}~\bibnamefont {Berg}}, \bibinfo {author}
  {\bibfnamefont {N.~H.}\ \bibnamefont {Lindner}}, \ and\ \bibinfo {author}
  {\bibfnamefont {M.~S.}\ \bibnamefont {Rudner}},\ }\href {\doibase
  10.1103/PhysRevB.99.195133} {\bibfield  {journal} {\bibinfo  {journal} {Phys.
  Rev. B}\ }\textbf {\bibinfo {volume} {99}},\ \bibinfo {pages} {195133}
  (\bibinfo {year} {2019})}\BibitemShut {NoStop}%
\bibitem [{\citenamefont {Po}\ \emph {et~al.}(2016)\citenamefont {Po},
  \citenamefont {Fidkowski}, \citenamefont {Morimoto}, \citenamefont {Potter},\
  and\ \citenamefont {Vishwanath}}]{PhysRevX.6.041070}%
  \BibitemOpen
  \bibfield  {author} {\bibinfo {author} {\bibfnamefont {H.~C.}\ \bibnamefont
  {Po}}, \bibinfo {author} {\bibfnamefont {L.}~\bibnamefont {Fidkowski}},
  \bibinfo {author} {\bibfnamefont {T.}~\bibnamefont {Morimoto}}, \bibinfo
  {author} {\bibfnamefont {A.~C.}\ \bibnamefont {Potter}}, \ and\ \bibinfo
  {author} {\bibfnamefont {A.}~\bibnamefont {Vishwanath}},\ }\href {\doibase
  10.1103/PhysRevX.6.041070} {\bibfield  {journal} {\bibinfo  {journal} {Phys.
  Rev. X}\ }\textbf {\bibinfo {volume} {6}},\ \bibinfo {pages} {041070}
  (\bibinfo {year} {2016})}\BibitemShut {NoStop}%
\bibitem [{\citenamefont {Lindner}\ \emph {et~al.}(2017)\citenamefont
  {Lindner}, \citenamefont {Berg},\ and\ \citenamefont
  {Rudner}}]{PhysRevX.7.011018}%
  \BibitemOpen
  \bibfield  {author} {\bibinfo {author} {\bibfnamefont {N.~H.}\ \bibnamefont
  {Lindner}}, \bibinfo {author} {\bibfnamefont {E.}~\bibnamefont {Berg}}, \
  and\ \bibinfo {author} {\bibfnamefont {M.~S.}\ \bibnamefont {Rudner}},\
  }\href {\doibase 10.1103/PhysRevX.7.011018} {\bibfield  {journal} {\bibinfo
  {journal} {Phys. Rev. X}\ }\textbf {\bibinfo {volume} {7}},\ \bibinfo {pages}
  {011018} (\bibinfo {year} {2017})}\BibitemShut {NoStop}%
\bibitem [{\citenamefont {Berges}\ \emph {et~al.}(2004)\citenamefont {Berges},
  \citenamefont {Bors\'anyi},\ and\ \citenamefont
  {Wetterich}}]{PhysRevLett.93.142002}%
  \BibitemOpen
  \bibfield  {author} {\bibinfo {author} {\bibfnamefont {J.}~\bibnamefont
  {Berges}}, \bibinfo {author} {\bibfnamefont {S.}~\bibnamefont {Bors\'anyi}},
  \ and\ \bibinfo {author} {\bibfnamefont {C.}~\bibnamefont {Wetterich}},\
  }\href {\doibase 10.1103/PhysRevLett.93.142002} {\bibfield  {journal}
  {\bibinfo  {journal} {Phys. Rev. Lett.}\ }\textbf {\bibinfo {volume} {93}},\
  \bibinfo {pages} {142002} (\bibinfo {year} {2004})}\BibitemShut {NoStop}%
\bibitem [{\citenamefont {Kinoshita}\ \emph {et~al.}(2006)\citenamefont
  {Kinoshita}, \citenamefont {Wenger},\ and\ \citenamefont
  {Weiss}}]{kinoshita2006quantum}%
  \BibitemOpen
  \bibfield  {author} {\bibinfo {author} {\bibfnamefont {T.}~\bibnamefont
  {Kinoshita}}, \bibinfo {author} {\bibfnamefont {T.}~\bibnamefont {Wenger}}, \
  and\ \bibinfo {author} {\bibfnamefont {D.~S.}\ \bibnamefont {Weiss}},\ }\href
  {\doibase 10.1038/nature04693} {\bibfield  {journal} {\bibinfo  {journal}
  {Nature}\ }\textbf {\bibinfo {volume} {440}},\ \bibinfo {pages} {900}
  (\bibinfo {year} {2006})}\BibitemShut {NoStop}%
\bibitem [{\citenamefont {Moeckel}\ and\ \citenamefont
  {Kehrein}(2008)}]{moeckel2008interaction}%
  \BibitemOpen
  \bibfield  {author} {\bibinfo {author} {\bibfnamefont {M.}~\bibnamefont
  {Moeckel}}\ and\ \bibinfo {author} {\bibfnamefont {S.}~\bibnamefont
  {Kehrein}},\ }\href {\doibase 10.1103/PhysRevLett.100.175702} {\bibfield
  {journal} {\bibinfo  {journal} {Phys. Rev. Lett.}\ }\textbf {\bibinfo
  {volume} {100}},\ \bibinfo {pages} {175702} (\bibinfo {year}
  {2008})}\BibitemShut {NoStop}%
\bibitem [{\citenamefont {Eckstein}\ \emph {et~al.}(2009)\citenamefont
  {Eckstein}, \citenamefont {Kollar},\ and\ \citenamefont
  {Werner}}]{eckstein2009thermalization}%
  \BibitemOpen
  \bibfield  {author} {\bibinfo {author} {\bibfnamefont {M.}~\bibnamefont
  {Eckstein}}, \bibinfo {author} {\bibfnamefont {M.}~\bibnamefont {Kollar}}, \
  and\ \bibinfo {author} {\bibfnamefont {P.}~\bibnamefont {Werner}},\ }\href
  {\doibase 10.1103/PhysRevLett.103.056403} {\bibfield  {journal} {\bibinfo
  {journal} {Phys. Rev. Lett.}\ }\textbf {\bibinfo {volume} {103}},\ \bibinfo
  {pages} {056403} (\bibinfo {year} {2009})}\BibitemShut {NoStop}%
\bibitem [{\citenamefont {Gring}\ \emph {et~al.}(2012)\citenamefont {Gring},
  \citenamefont {Kuhnert}, \citenamefont {Langen}, \citenamefont {Kitagawa},
  \citenamefont {Rauer}, \citenamefont {Schreitl}, \citenamefont {Mazets},
  \citenamefont {Smith}, \citenamefont {Demler},\ and\ \citenamefont
  {Schmiedmayer}}]{gring2012relaxation}%
  \BibitemOpen
  \bibfield  {author} {\bibinfo {author} {\bibfnamefont {M.}~\bibnamefont
  {Gring}}, \bibinfo {author} {\bibfnamefont {M.}~\bibnamefont {Kuhnert}},
  \bibinfo {author} {\bibfnamefont {T.}~\bibnamefont {Langen}}, \bibinfo
  {author} {\bibfnamefont {T.}~\bibnamefont {Kitagawa}}, \bibinfo {author}
  {\bibfnamefont {B.}~\bibnamefont {Rauer}}, \bibinfo {author} {\bibfnamefont
  {M.}~\bibnamefont {Schreitl}}, \bibinfo {author} {\bibfnamefont
  {I.}~\bibnamefont {Mazets}}, \bibinfo {author} {\bibfnamefont {D.~A.}\
  \bibnamefont {Smith}}, \bibinfo {author} {\bibfnamefont {E.}~\bibnamefont
  {Demler}}, \ and\ \bibinfo {author} {\bibfnamefont {J.}~\bibnamefont
  {Schmiedmayer}},\ }\href {\doibase 10.1126/science.1224953} {\bibfield
  {journal} {\bibinfo  {journal} {Science}\ }\textbf {\bibinfo {volume}
  {337}},\ \bibinfo {pages} {1318–1322} (\bibinfo {year} {2012})}\BibitemShut
  {NoStop}%
\bibitem [{\citenamefont {Bertini}\ \emph {et~al.}(2015)\citenamefont
  {Bertini}, \citenamefont {Essler}, \citenamefont {Groha},\ and\ \citenamefont
  {Robinson}}]{bertini2015prethermalization}%
  \BibitemOpen
  \bibfield  {author} {\bibinfo {author} {\bibfnamefont {B.}~\bibnamefont
  {Bertini}}, \bibinfo {author} {\bibfnamefont {F.~H.~L.}\ \bibnamefont
  {Essler}}, \bibinfo {author} {\bibfnamefont {S.}~\bibnamefont {Groha}}, \
  and\ \bibinfo {author} {\bibfnamefont {N.~J.}\ \bibnamefont {Robinson}},\
  }\href {\doibase 10.1103/PhysRevLett.115.180601} {\bibfield  {journal}
  {\bibinfo  {journal} {Phys. Rev. Lett.}\ }\textbf {\bibinfo {volume} {115}},\
  \bibinfo {pages} {180601} (\bibinfo {year} {2015})}\BibitemShut {NoStop}%
\bibitem [{\citenamefont {Mori}\ \emph {et~al.}(2016)\citenamefont {Mori},
  \citenamefont {Kuwahara},\ and\ \citenamefont
  {Saito}}]{PhysRevLett.116.120401}%
  \BibitemOpen
  \bibfield  {author} {\bibinfo {author} {\bibfnamefont {T.}~\bibnamefont
  {Mori}}, \bibinfo {author} {\bibfnamefont {T.}~\bibnamefont {Kuwahara}}, \
  and\ \bibinfo {author} {\bibfnamefont {K.}~\bibnamefont {Saito}},\ }\href
  {\doibase 10.1103/PhysRevLett.116.120401} {\bibfield  {journal} {\bibinfo
  {journal} {Phys. Rev. Lett.}\ }\textbf {\bibinfo {volume} {116}},\ \bibinfo
  {pages} {120401} (\bibinfo {year} {2016})}\BibitemShut {NoStop}%
\bibitem [{\citenamefont {Kuwahara}\ \emph {et~al.}(2016)\citenamefont
  {Kuwahara}, \citenamefont {Mori},\ and\ \citenamefont
  {Saito}}]{KUWAHARA201696}%
  \BibitemOpen
  \bibfield  {author} {\bibinfo {author} {\bibfnamefont {T.}~\bibnamefont
  {Kuwahara}}, \bibinfo {author} {\bibfnamefont {T.}~\bibnamefont {Mori}}, \
  and\ \bibinfo {author} {\bibfnamefont {K.}~\bibnamefont {Saito}},\ }\href
  {\doibase https://doi.org/10.1016/j.aop.2016.01.012} {\bibfield  {journal}
  {\bibinfo  {journal} {Annals of Physics}\ }\textbf {\bibinfo {volume}
  {367}},\ \bibinfo {pages} {96 } (\bibinfo {year} {2016})}\BibitemShut
  {NoStop}%
\bibitem [{\citenamefont {Abanin}\ \emph
  {et~al.}(2017{\natexlab{a}})\citenamefont {Abanin}, \citenamefont {De~Roeck},
  \citenamefont {Ho},\ and\ \citenamefont {Huveneers}}]{Abanin2017}%
  \BibitemOpen
  \bibfield  {author} {\bibinfo {author} {\bibfnamefont {D.}~\bibnamefont
  {Abanin}}, \bibinfo {author} {\bibfnamefont {W.}~\bibnamefont {De~Roeck}},
  \bibinfo {author} {\bibfnamefont {W.~W.}\ \bibnamefont {Ho}}, \ and\ \bibinfo
  {author} {\bibfnamefont {F.}~\bibnamefont {Huveneers}},\ }\href {\doibase
  10.1007/s00220-017-2930-x} {\bibfield  {journal} {\bibinfo  {journal}
  {Communications in Mathematical Physics}\ }\textbf {\bibinfo {volume}
  {354}},\ \bibinfo {pages} {809} (\bibinfo {year}
  {2017}{\natexlab{a}})}\BibitemShut {NoStop}%
\bibitem [{\citenamefont {Abanin}\ \emph
  {et~al.}(2017{\natexlab{b}})\citenamefont {Abanin}, \citenamefont {De~Roeck},
  \citenamefont {Ho},\ and\ \citenamefont {Huveneers}}]{PhysRevB.95.014112}%
  \BibitemOpen
  \bibfield  {author} {\bibinfo {author} {\bibfnamefont {D.~A.}\ \bibnamefont
  {Abanin}}, \bibinfo {author} {\bibfnamefont {W.}~\bibnamefont {De~Roeck}},
  \bibinfo {author} {\bibfnamefont {W.~W.}\ \bibnamefont {Ho}}, \ and\ \bibinfo
  {author} {\bibfnamefont {F.}~\bibnamefont {Huveneers}},\ }\href {\doibase
  10.1103/PhysRevB.95.014112} {\bibfield  {journal} {\bibinfo  {journal} {Phys.
  Rev. B}\ }\textbf {\bibinfo {volume} {95}},\ \bibinfo {pages} {014112}
  (\bibinfo {year} {2017}{\natexlab{b}})}\BibitemShut {NoStop}%
\bibitem [{\citenamefont {Howell}\ \emph {et~al.}(2019)\citenamefont {Howell},
  \citenamefont {Weinberg}, \citenamefont {Sels}, \citenamefont {Polkovnikov},\
  and\ \citenamefont {Bukov}}]{howell2019asymptotic}%
  \BibitemOpen
  \bibfield  {author} {\bibinfo {author} {\bibfnamefont {O.}~\bibnamefont
  {Howell}}, \bibinfo {author} {\bibfnamefont {P.}~\bibnamefont {Weinberg}},
  \bibinfo {author} {\bibfnamefont {D.}~\bibnamefont {Sels}}, \bibinfo {author}
  {\bibfnamefont {A.}~\bibnamefont {Polkovnikov}}, \ and\ \bibinfo {author}
  {\bibfnamefont {M.}~\bibnamefont {Bukov}},\ }\href {\doibase
  10.1103/PhysRevLett.122.010602} {\bibfield  {journal} {\bibinfo  {journal}
  {Phys. Rev. Lett.}\ }\textbf {\bibinfo {volume} {122}},\ \bibinfo {pages}
  {010602} (\bibinfo {year} {2019})}\BibitemShut {NoStop}%
\bibitem [{\citenamefont {Mori}(2018)}]{mori2018floquet}%
  \BibitemOpen
  \bibfield  {author} {\bibinfo {author} {\bibfnamefont {T.}~\bibnamefont
  {Mori}},\ }\href {\doibase 10.1103/PhysRevB.98.104303} {\bibfield  {journal}
  {\bibinfo  {journal} {Phys. Rev. B}\ }\textbf {\bibinfo {volume} {98}},\
  \bibinfo {pages} {104303} (\bibinfo {year} {2018})}\BibitemShut {NoStop}%
\bibitem [{\citenamefont {Rajak}\ \emph {et~al.}(2018)\citenamefont {Rajak},
  \citenamefont {Citro},\ and\ \citenamefont
  {Dalla~Torre}}]{rajak2018stability}%
  \BibitemOpen
  \bibfield  {author} {\bibinfo {author} {\bibfnamefont {A.}~\bibnamefont
  {Rajak}}, \bibinfo {author} {\bibfnamefont {R.}~\bibnamefont {Citro}}, \ and\
  \bibinfo {author} {\bibfnamefont {E.~G.}\ \bibnamefont {Dalla~Torre}},\
  }\href {\doibase 10.1088/1751-8121/aae294} {\bibfield  {journal} {\bibinfo
  {journal} {Journal of Physics A: Mathematical and Theoretical}\ }\textbf
  {\bibinfo {volume} {51}},\ \bibinfo {pages} {465001} (\bibinfo {year}
  {2018})}\BibitemShut {NoStop}%
\bibitem [{\citenamefont {Mallayya}\ \emph {et~al.}(2019)\citenamefont
  {Mallayya}, \citenamefont {Rigol},\ and\ \citenamefont
  {De~Roeck}}]{PhysRevX.9.021027}%
  \BibitemOpen
  \bibfield  {author} {\bibinfo {author} {\bibfnamefont {K.}~\bibnamefont
  {Mallayya}}, \bibinfo {author} {\bibfnamefont {M.}~\bibnamefont {Rigol}}, \
  and\ \bibinfo {author} {\bibfnamefont {W.}~\bibnamefont {De~Roeck}},\ }\href
  {\doibase 10.1103/PhysRevX.9.021027} {\bibfield  {journal} {\bibinfo
  {journal} {Phys. Rev. X}\ }\textbf {\bibinfo {volume} {9}},\ \bibinfo {pages}
  {021027} (\bibinfo {year} {2019})}\BibitemShut {NoStop}%
\bibitem [{\citenamefont {Reimann}\ and\ \citenamefont
  {Dabelow}(2019)}]{reimann2019typicality}%
  \BibitemOpen
  \bibfield  {author} {\bibinfo {author} {\bibfnamefont {P.}~\bibnamefont
  {Reimann}}\ and\ \bibinfo {author} {\bibfnamefont {L.}~\bibnamefont
  {Dabelow}},\ }\href {\doibase 10.1103/PhysRevLett.122.080603} {\bibfield
  {journal} {\bibinfo  {journal} {Phys. Rev. Lett.}\ }\textbf {\bibinfo
  {volume} {122}},\ \bibinfo {pages} {080603} (\bibinfo {year}
  {2019})}\BibitemShut {NoStop}%
\bibitem [{\citenamefont {Huveneers}\ and\ \citenamefont
  {Lukkarinen}(2020)}]{huveneers2020prethermalization}%
  \BibitemOpen
  \bibfield  {author} {\bibinfo {author} {\bibfnamefont {F.}~\bibnamefont
  {Huveneers}}\ and\ \bibinfo {author} {\bibfnamefont {J.}~\bibnamefont
  {Lukkarinen}},\ }\href {\doibase 10.1103/PhysRevResearch.2.022034} {\bibfield
   {journal} {\bibinfo  {journal} {Phys. Rev. Research}\ }\textbf {\bibinfo
  {volume} {2}},\ \bibinfo {pages} {022034} (\bibinfo {year}
  {2020})}\BibitemShut {NoStop}%
\bibitem [{\citenamefont {Rubio-Abadal}\ \emph {et~al.}(2020)\citenamefont
  {Rubio-Abadal}, \citenamefont {Ippoliti}, \citenamefont {Hollerith},
  \citenamefont {Wei}, \citenamefont {Rui}, \citenamefont {Sondhi},
  \citenamefont {Khemani}, \citenamefont {Gross},\ and\ \citenamefont
  {Bloch}}]{rubio2020floquet}%
  \BibitemOpen
  \bibfield  {author} {\bibinfo {author} {\bibfnamefont {A.}~\bibnamefont
  {Rubio-Abadal}}, \bibinfo {author} {\bibfnamefont {M.}~\bibnamefont
  {Ippoliti}}, \bibinfo {author} {\bibfnamefont {S.}~\bibnamefont {Hollerith}},
  \bibinfo {author} {\bibfnamefont {D.}~\bibnamefont {Wei}}, \bibinfo {author}
  {\bibfnamefont {J.}~\bibnamefont {Rui}}, \bibinfo {author} {\bibfnamefont
  {S.~L.}\ \bibnamefont {Sondhi}}, \bibinfo {author} {\bibfnamefont
  {V.}~\bibnamefont {Khemani}}, \bibinfo {author} {\bibfnamefont
  {C.}~\bibnamefont {Gross}}, \ and\ \bibinfo {author} {\bibfnamefont
  {I.}~\bibnamefont {Bloch}},\ }\href {\doibase 10.1103/PhysRevX.10.021044}
  {\bibfield  {journal} {\bibinfo  {journal} {Phys. Rev. X}\ }\textbf {\bibinfo
  {volume} {10}},\ \bibinfo {pages} {021044} (\bibinfo {year}
  {2020})}\BibitemShut {NoStop}%
\bibitem [{\citenamefont {Lazarides}\ \emph {et~al.}(2014)\citenamefont
  {Lazarides}, \citenamefont {Das},\ and\ \citenamefont
  {Moessner}}]{lazarides_das_14}%
  \BibitemOpen
  \bibfield  {author} {\bibinfo {author} {\bibfnamefont {A.}~\bibnamefont
  {Lazarides}}, \bibinfo {author} {\bibfnamefont {A.}~\bibnamefont {Das}}, \
  and\ \bibinfo {author} {\bibfnamefont {R.}~\bibnamefont {Moessner}},\ }\href
  {\doibase 10.1103/PhysRevE.90.012110} {\bibfield  {journal} {\bibinfo
  {journal} {Phys. Rev. E}\ }\textbf {\bibinfo {volume} {90}},\ \bibinfo
  {pages} {012110} (\bibinfo {year} {2014})}\BibitemShut {NoStop}%
\bibitem [{\citenamefont {D'Alessio}\ and\ \citenamefont
  {Rigol}(2014)}]{dalessio_rigol_14}%
  \BibitemOpen
  \bibfield  {author} {\bibinfo {author} {\bibfnamefont {L.}~\bibnamefont
  {D'Alessio}}\ and\ \bibinfo {author} {\bibfnamefont {M.}~\bibnamefont
  {Rigol}},\ }\href {\doibase 10.1103/PhysRevX.4.041048} {\bibfield  {journal}
  {\bibinfo  {journal} {Phys. Rev. X}\ }\textbf {\bibinfo {volume} {4}},\
  \bibinfo {pages} {041048} (\bibinfo {year} {2014})}\BibitemShut {NoStop}%
\bibitem [{\citenamefont {Bukov}\ \emph {et~al.}(2015)\citenamefont {Bukov},
  \citenamefont {D’Alessio},\ and\ \citenamefont
  {Polkovnikov}}]{bukov2015universal}%
  \BibitemOpen
  \bibfield  {author} {\bibinfo {author} {\bibfnamefont {M.}~\bibnamefont
  {Bukov}}, \bibinfo {author} {\bibfnamefont {L.}~\bibnamefont {D’Alessio}},
  \ and\ \bibinfo {author} {\bibfnamefont {A.}~\bibnamefont {Polkovnikov}},\
  }\href {\doibase 10.1080/00018732.2015.1055918} {\bibfield  {journal}
  {\bibinfo  {journal} {Advances in Physics}\ }\textbf {\bibinfo {volume}
  {64}},\ \bibinfo {pages} {139–226} (\bibinfo {year} {2015})}\BibitemShut
  {NoStop}%
\bibitem [{\citenamefont {Else}\ \emph {et~al.}(2017)\citenamefont {Else},
  \citenamefont {Bauer},\ and\ \citenamefont {Nayak}}]{PhysRevX.7.011026}%
  \BibitemOpen
  \bibfield  {author} {\bibinfo {author} {\bibfnamefont {D.~V.}\ \bibnamefont
  {Else}}, \bibinfo {author} {\bibfnamefont {B.}~\bibnamefont {Bauer}}, \ and\
  \bibinfo {author} {\bibfnamefont {C.}~\bibnamefont {Nayak}},\ }\href
  {\doibase 10.1103/PhysRevX.7.011026} {\bibfield  {journal} {\bibinfo
  {journal} {Phys. Rev. X}\ }\textbf {\bibinfo {volume} {7}},\ \bibinfo {pages}
  {011026} (\bibinfo {year} {2017})}\BibitemShut {NoStop}%
\bibitem [{\citenamefont {Dumitrescu}\ \emph {et~al.}(2018)\citenamefont
  {Dumitrescu}, \citenamefont {Vasseur},\ and\ \citenamefont
  {Potter}}]{dumitrescu2018logarithmically}%
  \BibitemOpen
  \bibfield  {author} {\bibinfo {author} {\bibfnamefont {P.~T.}\ \bibnamefont
  {Dumitrescu}}, \bibinfo {author} {\bibfnamefont {R.}~\bibnamefont {Vasseur}},
  \ and\ \bibinfo {author} {\bibfnamefont {A.~C.}\ \bibnamefont {Potter}},\
  }\href {\doibase 10.1103/PhysRevLett.120.070602} {\bibfield  {journal}
  {\bibinfo  {journal} {Phys. Rev. Lett.}\ }\textbf {\bibinfo {volume} {120}},\
  \bibinfo {pages} {070602} (\bibinfo {year} {2018})}\BibitemShut {NoStop}%
\bibitem [{\citenamefont {Else}\ \emph {et~al.}(2020)\citenamefont {Else},
  \citenamefont {Ho},\ and\ \citenamefont {Dumitrescu}}]{PhysRevX.10.021032}%
  \BibitemOpen
  \bibfield  {author} {\bibinfo {author} {\bibfnamefont {D.~V.}\ \bibnamefont
  {Else}}, \bibinfo {author} {\bibfnamefont {W.~W.}\ \bibnamefont {Ho}}, \ and\
  \bibinfo {author} {\bibfnamefont {P.~T.}\ \bibnamefont {Dumitrescu}},\ }\href
  {\doibase 10.1103/PhysRevX.10.021032} {\bibfield  {journal} {\bibinfo
  {journal} {Phys. Rev. X}\ }\textbf {\bibinfo {volume} {10}},\ \bibinfo
  {pages} {021032} (\bibinfo {year} {2020})}\BibitemShut {NoStop}%
\bibitem [{\citenamefont {Blanes}\ \emph {et~al.}(2009)\citenamefont {Blanes},
  \citenamefont {Casas}, \citenamefont {Oteo},\ and\ \citenamefont
  {Ros}}]{Blanes_2009}%
  \BibitemOpen
  \bibfield  {author} {\bibinfo {author} {\bibfnamefont {S.}~\bibnamefont
  {Blanes}}, \bibinfo {author} {\bibfnamefont {F.}~\bibnamefont {Casas}},
  \bibinfo {author} {\bibfnamefont {J.}~\bibnamefont {Oteo}}, \ and\ \bibinfo
  {author} {\bibfnamefont {J.}~\bibnamefont {Ros}},\ }\href {\doibase
  10.1016/j.physrep.2008.11.001} {\bibfield  {journal} {\bibinfo  {journal}
  {Physics Reports}\ }\textbf {\bibinfo {volume} {470}},\ \bibinfo {pages}
  {151–238} (\bibinfo {year} {2009})}\BibitemShut {NoStop}%
\bibitem [{\citenamefont {Luitz}\ \emph {et~al.}(2020)\citenamefont {Luitz},
  \citenamefont {Moessner}, \citenamefont {Sondhi},\ and\ \citenamefont
  {Khemani}}]{PhysRevX.10.021046}%
  \BibitemOpen
  \bibfield  {author} {\bibinfo {author} {\bibfnamefont {D.~J.}\ \bibnamefont
  {Luitz}}, \bibinfo {author} {\bibfnamefont {R.}~\bibnamefont {Moessner}},
  \bibinfo {author} {\bibfnamefont {S.~L.}\ \bibnamefont {Sondhi}}, \ and\
  \bibinfo {author} {\bibfnamefont {V.}~\bibnamefont {Khemani}},\ }\href
  {\doibase 10.1103/PhysRevX.10.021046} {\bibfield  {journal} {\bibinfo
  {journal} {Phys. Rev. X}\ }\textbf {\bibinfo {volume} {10}},\ \bibinfo
  {pages} {021046} (\bibinfo {year} {2020})}\BibitemShut {NoStop}%
\bibitem [{\citenamefont {Gulden}\ \emph {et~al.}(2020)\citenamefont {Gulden},
  \citenamefont {Berg}, \citenamefont {Rudner},\ and\ \citenamefont
  {Lindner}}]{Gulden_2020}%
  \BibitemOpen
  \bibfield  {author} {\bibinfo {author} {\bibfnamefont {T.}~\bibnamefont
  {Gulden}}, \bibinfo {author} {\bibfnamefont {E.}~\bibnamefont {Berg}},
  \bibinfo {author} {\bibfnamefont {M.~S.}\ \bibnamefont {Rudner}}, \ and\
  \bibinfo {author} {\bibfnamefont {N.}~\bibnamefont {Lindner}},\ }\href
  {\doibase 10.21468/scipostphys.9.1.015} {\bibfield  {journal} {\bibinfo
  {journal} {SciPost Physics}\ }\textbf {\bibinfo {volume} {9}} (\bibinfo
  {year} {2020}),\ 10.21468/scipostphys.9.1.015}\BibitemShut {NoStop}%
\bibitem [{\citenamefont {Bravyi}\ \emph {et~al.}(2011)\citenamefont {Bravyi},
  \citenamefont {DiVincenzo},\ and\ \citenamefont {Loss}}]{Bravyi_2011}%
  \BibitemOpen
  \bibfield  {author} {\bibinfo {author} {\bibfnamefont {S.}~\bibnamefont
  {Bravyi}}, \bibinfo {author} {\bibfnamefont {D.~P.}\ \bibnamefont
  {DiVincenzo}}, \ and\ \bibinfo {author} {\bibfnamefont {D.}~\bibnamefont
  {Loss}},\ }\href {\doibase 10.1016/j.aop.2011.06.004} {\bibfield  {journal}
  {\bibinfo  {journal} {Annals of Physics}\ }\textbf {\bibinfo {volume}
  {326}},\ \bibinfo {pages} {2793–2826} (\bibinfo {year} {2011})}\BibitemShut
  {NoStop}%
\bibitem [{SM()}]{SM}%
  \BibitemOpen
  \href@noop {} {}\bibinfo {note} {See Supplemental Material online for proofs
  of theorems, extensions of results, and details on physical
  consequences.}\BibitemShut {Stop}%
\bibitem [{\citenamefont {Lieb}\ and\ \citenamefont
  {Robinson}(1972)}]{Lieb1972}%
  \BibitemOpen
  \bibfield  {author} {\bibinfo {author} {\bibfnamefont {E.~H.}\ \bibnamefont
  {Lieb}}\ and\ \bibinfo {author} {\bibfnamefont {D.~W.}\ \bibnamefont
  {Robinson}},\ }\href {\doibase 10.1007/BF01645779} {\bibfield  {journal}
  {\bibinfo  {journal} {Communications in Mathematical Physics}\ }\textbf
  {\bibinfo {volume} {28}},\ \bibinfo {pages} {251} (\bibinfo {year}
  {1972})}\BibitemShut {NoStop}%
\bibitem [{\citenamefont {Machado}\ \emph {et~al.}(2020)\citenamefont
  {Machado}, \citenamefont {Else}, \citenamefont {Kahanamoku-Meyer},
  \citenamefont {Nayak},\ and\ \citenamefont {Yao}}]{PhysRevX.10.011043}%
  \BibitemOpen
  \bibfield  {author} {\bibinfo {author} {\bibfnamefont {F.}~\bibnamefont
  {Machado}}, \bibinfo {author} {\bibfnamefont {D.~V.}\ \bibnamefont {Else}},
  \bibinfo {author} {\bibfnamefont {G.~D.}\ \bibnamefont {Kahanamoku-Meyer}},
  \bibinfo {author} {\bibfnamefont {C.}~\bibnamefont {Nayak}}, \ and\ \bibinfo
  {author} {\bibfnamefont {N.~Y.}\ \bibnamefont {Yao}},\ }\href {\doibase
  10.1103/PhysRevX.10.011043} {\bibfield  {journal} {\bibinfo  {journal} {Phys.
  Rev. X}\ }\textbf {\bibinfo {volume} {10}},\ \bibinfo {pages} {011043}
  (\bibinfo {year} {2020})}\BibitemShut {NoStop}%
\bibitem [{\citenamefont {Datta}\ \emph {et~al.}(1996)\citenamefont {Datta},
  \citenamefont {Fern{\'a}ndez},\ and\ \citenamefont
  {Fr{\"o}hlich}}]{datta1996low}%
  \BibitemOpen
  \bibfield  {author} {\bibinfo {author} {\bibfnamefont {N.}~\bibnamefont
  {Datta}}, \bibinfo {author} {\bibfnamefont {R.}~\bibnamefont
  {Fern{\'a}ndez}}, \ and\ \bibinfo {author} {\bibfnamefont {J.}~\bibnamefont
  {Fr{\"o}hlich}},\ }\href {\doibase 10.1007/BF02179651} {\bibfield  {journal}
  {\bibinfo  {journal} {Journal of Statistical Physics}\ }\textbf {\bibinfo
  {volume} {84}},\ \bibinfo {pages} {455} (\bibinfo {year} {1996})}\BibitemShut
  {NoStop}%
\bibitem [{\citenamefont {Benettin}\ \emph {et~al.}(1988)\citenamefont
  {Benettin}, \citenamefont {Fr{\"o}hlich},\ and\ \citenamefont
  {Giorgilli}}]{benettin1988nekhoroshev}%
  \BibitemOpen
  \bibfield  {author} {\bibinfo {author} {\bibfnamefont {G.}~\bibnamefont
  {Benettin}}, \bibinfo {author} {\bibfnamefont {J.}~\bibnamefont
  {Fr{\"o}hlich}}, \ and\ \bibinfo {author} {\bibfnamefont {A.}~\bibnamefont
  {Giorgilli}},\ }\href {\doibase 10.1007/BF01218262} {\bibfield  {journal}
  {\bibinfo  {journal} {Communications in Mathematical Physics}\ }\textbf
  {\bibinfo {volume} {119}},\ \bibinfo {pages} {95} (\bibinfo {year}
  {1988})}\BibitemShut {NoStop}%
\bibitem [{\citenamefont {Cuneo}\ \emph {et~al.}(2017)\citenamefont {Cuneo},
  \citenamefont {Eckmann},\ and\ \citenamefont {Wayne}}]{cuneo2017energy}%
  \BibitemOpen
  \bibfield  {author} {\bibinfo {author} {\bibfnamefont {N.}~\bibnamefont
  {Cuneo}}, \bibinfo {author} {\bibfnamefont {J.-P.}\ \bibnamefont {Eckmann}},
  \ and\ \bibinfo {author} {\bibfnamefont {C.~E.}\ \bibnamefont {Wayne}},\
  }\href {\doibase 10.1088/1361-6544/aa85d6} {\bibfield  {journal} {\bibinfo
  {journal} {Nonlinearity}\ }\textbf {\bibinfo {volume} {30}},\ \bibinfo
  {pages} {R81–R117} (\bibinfo {year} {2017})}\BibitemShut {NoStop}%
\bibitem [{\citenamefont {Giorgilli}\ \emph {et~al.}(2014)\citenamefont
  {Giorgilli}, \citenamefont {Paleari},\ and\ \citenamefont
  {Penati}}]{giorgilli2015extensive}%
  \BibitemOpen
  \bibfield  {author} {\bibinfo {author} {\bibfnamefont {A.}~\bibnamefont
  {Giorgilli}}, \bibinfo {author} {\bibfnamefont {S.}~\bibnamefont {Paleari}},
  \ and\ \bibinfo {author} {\bibfnamefont {T.}~\bibnamefont {Penati}},\ }\href
  {\doibase 10.1007/s00023-014-0335-3} {\bibfield  {journal} {\bibinfo
  {journal} {Annales Henri Poincaré}\ }\textbf {\bibinfo {volume} {16}},\
  \bibinfo {pages} {897–959} (\bibinfo {year} {2014})}\BibitemShut {NoStop}%
\bibitem [{\citenamefont {Fr{\"o}hlich}\ \emph {et~al.}(1986)\citenamefont
  {Fr{\"o}hlich}, \citenamefont {Spencer},\ and\ \citenamefont
  {Wayne}}]{frohlich1986localization}%
  \BibitemOpen
  \bibfield  {author} {\bibinfo {author} {\bibfnamefont {J.}~\bibnamefont
  {Fr{\"o}hlich}}, \bibinfo {author} {\bibfnamefont {T.}~\bibnamefont
  {Spencer}}, \ and\ \bibinfo {author} {\bibfnamefont {C.~E.}\ \bibnamefont
  {Wayne}},\ }\href {\doibase 10.1007/BF01127712} {\bibfield  {journal}
  {\bibinfo  {journal} {Journal of Statistical Physics}\ }\textbf {\bibinfo
  {volume} {42}},\ \bibinfo {pages} {247} (\bibinfo {year} {1986})}\BibitemShut
  {NoStop}%
\bibitem [{\citenamefont {Imbrie}(2016)}]{imbrie2016many}%
  \BibitemOpen
  \bibfield  {author} {\bibinfo {author} {\bibfnamefont {J.~Z.}\ \bibnamefont
  {Imbrie}},\ }\href {\doibase 10.1007/s10955-016-1508-x} {\bibfield  {journal}
  {\bibinfo  {journal} {Journal of Statistical Physics}\ }\textbf {\bibinfo
  {volume} {163}},\ \bibinfo {pages} {998–1048} (\bibinfo {year}
  {2016})}\BibitemShut {NoStop}%
\bibitem [{\citenamefont {Labuhn}\ \emph {et~al.}(2016)\citenamefont {Labuhn},
  \citenamefont {Barredo}, \citenamefont {Ravets}, \citenamefont
  {de~Léséleuc}, \citenamefont {Macrì}, \citenamefont {Lahaye},\ and\
  \citenamefont {Browaeys}}]{Labuhn_2016}%
  \BibitemOpen
  \bibfield  {author} {\bibinfo {author} {\bibfnamefont {H.}~\bibnamefont
  {Labuhn}}, \bibinfo {author} {\bibfnamefont {D.}~\bibnamefont {Barredo}},
  \bibinfo {author} {\bibfnamefont {S.}~\bibnamefont {Ravets}}, \bibinfo
  {author} {\bibfnamefont {S.}~\bibnamefont {de~Léséleuc}}, \bibinfo {author}
  {\bibfnamefont {T.}~\bibnamefont {Macrì}}, \bibinfo {author} {\bibfnamefont
  {T.}~\bibnamefont {Lahaye}}, \ and\ \bibinfo {author} {\bibfnamefont
  {A.}~\bibnamefont {Browaeys}},\ }\href {\doibase 10.1038/nature18274}
  {\bibfield  {journal} {\bibinfo  {journal} {Nature}\ }\textbf {\bibinfo
  {volume} {534}},\ \bibinfo {pages} {667–670} (\bibinfo {year}
  {2016})}\BibitemShut {NoStop}%
\bibitem [{\citenamefont {Bernien}\ \emph {et~al.}(2017)\citenamefont
  {Bernien}, \citenamefont {Schwartz}, \citenamefont {Keesling}, \citenamefont
  {Levine}, \citenamefont {Omran}, \citenamefont {Pichler}, \citenamefont
  {Choi}, \citenamefont {Zibrov}, \citenamefont {Endres}, \citenamefont
  {Greiner}, \citenamefont {Vuletic},\ and\ \citenamefont {Lukin}}]{Lukin1}%
  \BibitemOpen
  \bibfield  {author} {\bibinfo {author} {\bibfnamefont {H.}~\bibnamefont
  {Bernien}}, \bibinfo {author} {\bibfnamefont {S.}~\bibnamefont {Schwartz}},
  \bibinfo {author} {\bibfnamefont {A.}~\bibnamefont {Keesling}}, \bibinfo
  {author} {\bibfnamefont {H.}~\bibnamefont {Levine}}, \bibinfo {author}
  {\bibfnamefont {A.}~\bibnamefont {Omran}}, \bibinfo {author} {\bibfnamefont
  {H.}~\bibnamefont {Pichler}}, \bibinfo {author} {\bibfnamefont
  {S.}~\bibnamefont {Choi}}, \bibinfo {author} {\bibfnamefont {A.~S.}\
  \bibnamefont {Zibrov}}, \bibinfo {author} {\bibfnamefont {M.}~\bibnamefont
  {Endres}}, \bibinfo {author} {\bibfnamefont {M.}~\bibnamefont {Greiner}},
  \bibinfo {author} {\bibfnamefont {V.}~\bibnamefont {Vuletic}}, \ and\
  \bibinfo {author} {\bibfnamefont {M.~D.}\ \bibnamefont {Lukin}},\ }\href
  {https://doi.org/10.1038/nature24622} {\bibfield  {journal} {\bibinfo
  {journal} {Nature}\ }\textbf {\bibinfo {volume} {551}},\ \bibinfo {pages}
  {579 EP } (\bibinfo {year} {2017})},\ \bibinfo {note} {article}\BibitemShut
  {NoStop}%
\bibitem [{\citenamefont {Kim}\ \emph {et~al.}(2018)\citenamefont {Kim},
  \citenamefont {Park}, \citenamefont {Kim}, \citenamefont {Sim},\ and\
  \citenamefont {Ahn}}]{PhysRevLett.120.180502}%
  \BibitemOpen
  \bibfield  {author} {\bibinfo {author} {\bibfnamefont {H.}~\bibnamefont
  {Kim}}, \bibinfo {author} {\bibfnamefont {Y.}~\bibnamefont {Park}}, \bibinfo
  {author} {\bibfnamefont {K.}~\bibnamefont {Kim}}, \bibinfo {author}
  {\bibfnamefont {H.-S.}\ \bibnamefont {Sim}}, \ and\ \bibinfo {author}
  {\bibfnamefont {J.}~\bibnamefont {Ahn}},\ }\href {\doibase
  10.1103/PhysRevLett.120.180502} {\bibfield  {journal} {\bibinfo  {journal}
  {Phys. Rev. Lett.}\ }\textbf {\bibinfo {volume} {120}},\ \bibinfo {pages}
  {180502} (\bibinfo {year} {2018})}\BibitemShut {NoStop}%
\bibitem [{\citenamefont {Browaeys}\ and\ \citenamefont
  {Lahaye}(2020)}]{Browaeys_2020}%
  \BibitemOpen
  \bibfield  {author} {\bibinfo {author} {\bibfnamefont {A.}~\bibnamefont
  {Browaeys}}\ and\ \bibinfo {author} {\bibfnamefont {T.}~\bibnamefont
  {Lahaye}},\ }\href {\doibase 10.1038/s41567-019-0733-z} {\bibfield  {journal}
  {\bibinfo  {journal} {Nature Physics}\ }\textbf {\bibinfo {volume} {16}},\
  \bibinfo {pages} {132–142} (\bibinfo {year} {2020})}\BibitemShut {NoStop}%
\bibitem [{\citenamefont {Roeck}\ and\ \citenamefont
  {Verreet}(2019)}]{deroeck2019slow}%
  \BibitemOpen
  \bibfield  {author} {\bibinfo {author} {\bibfnamefont {W.~D.}\ \bibnamefont
  {Roeck}}\ and\ \bibinfo {author} {\bibfnamefont {V.}~\bibnamefont
  {Verreet}},\ }\href@noop {} {\enquote {\bibinfo {title} {Very slow heating
  for weakly driven quantum many-body systems},}\ } (\bibinfo {year} {2019}),\
  \Eprint {http://arxiv.org/abs/1911.01998} {arXiv:1911.01998
  [cond-mat.stat-mech]} \BibitemShut {NoStop}%
\end{thebibliography}%


\begin{thebibliography}{4}%
\makeatletter
\providecommand \@ifxundefined [1]{%
 \@ifx{#1\undefined}
}%
\providecommand \@ifnum [1]{%
 \ifnum #1\expandafter \@firstoftwo
 \else \expandafter \@secondoftwo
 \fi
}%
\providecommand \@ifx [1]{%
 \ifx #1\expandafter \@firstoftwo
 \else \expandafter \@secondoftwo
 \fi
}%
\providecommand \natexlab [1]{#1}%
\providecommand \enquote  [1]{``#1''}%
\providecommand \bibnamefont  [1]{#1}%
\providecommand \bibfnamefont [1]{#1}%
\providecommand \citenamefont [1]{#1}%
\providecommand \href@noop [0]{\@secondoftwo}%
\providecommand \href [0]{\begingroup \@sanitize@url \@href}%
\providecommand \@href[1]{\@@startlink{#1}\@@href}%
\providecommand \@@href[1]{\endgroup#1\@@endlink}%
\providecommand \@sanitize@url [0]{\catcode `\\12\catcode `\$12\catcode
  `\&12\catcode `\#12\catcode `\^12\catcode `\_12\catcode `\%12\relax}%
\providecommand \@@startlink[1]{}%
\providecommand \@@endlink[0]{}%
\providecommand \url  [0]{\begingroup\@sanitize@url \@url }%
\providecommand \@url [1]{\endgroup\@href {#1}{\urlprefix }}%
\providecommand \urlprefix  [0]{URL }%
\providecommand \Eprint [0]{\href }%
\providecommand \doibase [0]{http://dx.doi.org/}%
\providecommand \selectlanguage [0]{\@gobble}%
\providecommand \bibinfo  [0]{\@secondoftwo}%
\providecommand \bibfield  [0]{\@secondoftwo}%
\providecommand \translation [1]{[#1]}%
\providecommand \BibitemOpen [0]{}%
\providecommand \bibitemStop [0]{}%
\providecommand \bibitemNoStop [0]{.\EOS\space}%
\providecommand \EOS [0]{\spacefactor3000\relax}%
\providecommand \BibitemShut  [1]{\csname bibitem#1\endcsname}%
\let\auto@bib@innerbib\@empty
\bibitem [{\citenamefont {Else}\ \emph {et~al.}(2017)\citenamefont {Else},
  \citenamefont {Bauer},\ and\ \citenamefont {Nayak}}]{PhysRevX.7.011026}%
  \BibitemOpen
  \bibfield  {author} {\bibinfo {author} {\bibfnamefont {D.~V.}\ \bibnamefont
  {Else}}, \bibinfo {author} {\bibfnamefont {B.}~\bibnamefont {Bauer}}, \ and\
  \bibinfo {author} {\bibfnamefont {C.}~\bibnamefont {Nayak}},\ }\href
  {\doibase 10.1103/PhysRevX.7.011026} {\bibfield  {journal} {\bibinfo
  {journal} {Phys. Rev. X}\ }\textbf {\bibinfo {volume} {7}},\ \bibinfo {pages}
  {011026} (\bibinfo {year} {2017})}\BibitemShut {NoStop}%
\bibitem [{\citenamefont {Else}\ \emph {et~al.}(2020)\citenamefont {Else},
  \citenamefont {Ho},\ and\ \citenamefont {Dumitrescu}}]{PhysRevX.10.021032}%
  \BibitemOpen
  \bibfield  {author} {\bibinfo {author} {\bibfnamefont {D.~V.}\ \bibnamefont
  {Else}}, \bibinfo {author} {\bibfnamefont {W.~W.}\ \bibnamefont {Ho}}, \ and\
  \bibinfo {author} {\bibfnamefont {P.~T.}\ \bibnamefont {Dumitrescu}},\ }\href
  {\doibase 10.1103/PhysRevX.10.021032} {\bibfield  {journal} {\bibinfo
  {journal} {Phys. Rev. X}\ }\textbf {\bibinfo {volume} {10}},\ \bibinfo
  {pages} {021032} (\bibinfo {year} {2020})}\BibitemShut {NoStop}%
\bibitem [{\citenamefont {Abanin}\ \emph {et~al.}(2017)\citenamefont {Abanin},
  \citenamefont {De~Roeck}, \citenamefont {Ho},\ and\ \citenamefont
  {Huveneers}}]{Abanin2017}%
  \BibitemOpen
  \bibfield  {author} {\bibinfo {author} {\bibfnamefont {D.}~\bibnamefont
  {Abanin}}, \bibinfo {author} {\bibfnamefont {W.}~\bibnamefont {De~Roeck}},
  \bibinfo {author} {\bibfnamefont {W.~W.}\ \bibnamefont {Ho}}, \ and\ \bibinfo
  {author} {\bibfnamefont {F.}~\bibnamefont {Huveneers}},\ }\href {\doibase
  10.1007/s00220-017-2930-x} {\bibfield  {journal} {\bibinfo  {journal}
  {Communications in Mathematical Physics}\ }\textbf {\bibinfo {volume}
  {354}},\ \bibinfo {pages} {809} (\bibinfo {year} {2017})}\BibitemShut
  {NoStop}%
\bibitem [{\citenamefont {Lieb}\ and\ \citenamefont
  {Robinson}(1972)}]{Lieb1972}%
  \BibitemOpen
  \bibfield  {author} {\bibinfo {author} {\bibfnamefont {E.~H.}\ \bibnamefont
  {Lieb}}\ and\ \bibinfo {author} {\bibfnamefont {D.~W.}\ \bibnamefont
  {Robinson}},\ }\href {\doibase 10.1007/BF01645779} {\bibfield  {journal}
  {\bibinfo  {journal} {Communications in Mathematical Physics}\ }\textbf
  {\bibinfo {volume} {28}},\ \bibinfo {pages} {251} (\bibinfo {year}
  {1972})}\BibitemShut {NoStop}%
\end{thebibliography}%

\end{document}


\title{
Supplemental Material for: \\
A Rigorous Theory of Prethermalization without Temperature
}
\author{Wen Wei Ho}
\affiliation{Department of Physics, Stanford University, Stanford, CA 94305, USA } 

\author{Wojciech De Roeck}
\affiliation{Instituut Theoretische Fysica, KU Leuven, 3001 Leuven, Belgium}

\date{\today}
\maketitle


In this Supplemental Material, we provide  
(i) the proofs of Theorems 1 and 2 of the main text, corollaries on physical consequences such as long-lived charge conservation, and (ii) extensions of the theorems.

\section{Proofs of Theorems 1 and 2}
\subsection{Mathematical setting}
We consider a large but finite graph $\Lambda$, equipped with the graph distance. The vertices $i$ of this graph are our `sites'.  
We assume there is a finite Hilbert space $\mathbb{C}^d$ attached to each site $i\in \Lambda$ and we take $d$ to be fixed; the total  Hilbert space $\mathcal{H}$ is hence $(\mathbb{C}^d)^{\otimes_\Lambda}$.
We say that an operator $O = O_S$   in $\mathcal{B}\equiv\mathcal{B}(\mathcal{H})$ (the space of bounded operators on $\mathcal{H}$) is supported in a set $S$ if it is of the form $ O_S \otimes \mathbb{I}_{S^c}$ ($\mathbb{I}$: identity operator, $S^c: $ complement of $S$), with a slight abuse of notation.   This is the setting for quantum spin systems. One can also consider lattice fermions, if one makes some modifications in the definition to deal with the fact that the fermionic space is not naturally given as a tensor product (due to anticommutation relations). 
For operators $O \in \mathcal{B}$, we use the standard operator norm $\| O \|=\sup_{|\psi\rangle \in\mathcal{H}, \langle \psi |\psi\rangle \neq 0}  \langle \psi |O^\dagger O |\psi\rangle/\langle \psi|\psi\rangle $.
Also,   given an operator $A$, we will freely use the notation $\adjoint_A$ to denote the superoperator acting on $\mathcal{B}$ as 
$\adjoint_A(B)=[A,B].
$

 \subsubsection{The `number' operator $N$}

 The operator $N$ plays a central role in our analysis. We assume it is given as a sum of local terms $N=\sum_{S \subset \Lambda}N_{S}$ satisfying the following conditions:
\begin{enumerate}
\item All local terms mutually commute: $[N_S, N_{S'}]$. 
\item All of the $N_S$ have integer spectrum.
\item  There is a fixed range $R$ such that $N_S=0$ whenever $\text{diam}(S)>R$ (`diam' stands for diameter as defined in any metric space).
\end{enumerate}

With these definitions in hand, we need to refine the notion of support of operators, following \cite{PhysRevX.7.011026}.
 We say that $O \in \mathcal{B}$ is `strongly supported' in $S$ if
 $O$ is supported in $S$ and, for any $S' \not\subset S$ we have $[O,N_{S'}]=0$.  
Here are the important consequences:
\begin{enumerate}
\item For any function $f$, if $O$ is strongly supported in $S$, then 
$ f( \adjoint_N) O $ is strongly supported in $S$. 
\item If $A,B$ are strongly supported in $S_A,S_B$, then  $[A,B]$ is strongly supported in $S_A\cup S_B$. 
\end{enumerate}
We write $\mathcal{B}_S \subset \mathcal{B}$ for the algebra of operators strongly supported in $S$.

\subsubsection{Colored potential and norm}
We will manipulate operators that are not only sums of local terms (on the graph $\Lambda$), but also parameterized by angles  $\vec\theta \in [0,2\pi)^m$ ($m = 1$: Floquet; $m \geq 2$: quasiperiodic). To that end we introduce 
%
%
the notion of a `colored potential' $\Phi$, as was done in \cite{PhysRevX.10.021032}.  
This is a function 
\begin{align}
   2^\Lambda\times \mathbb{Z}^m \to \mathcal{B}: \bm{Z} := (Z,\vec{n}) \mapsto \Phi_{Z,\vec{n}}
\end{align}
  such that $\Phi_{Z,\vec{n}}\in \mathcal{B}_Z$ (i.e.~it has strong support in the set $Z$) and that $\Phi_{Z,\vec{n}}=0$ unless $Z$ is a connected set. 
 We define a weighted norm as
\begin{align}\label{def: norm}
\| \Phi \|_\kappa = \sup_x \sum_{\bm{Z} \ni x} e^{\kappa |\bm{Z}|} \| \Phi_{\bm{Z}} \|
\end{align}
for any $\kappa > 0$ where $x \in \bm{Z}$ iff $x \in Z$; also $|\bm{Z}| = |Z| + |\vec{n}|$.
 
We will also need the $\cup$ operation acting on colored sets as $(Z_1,\vec{n}_1) \cup (Z_2,\vec{n}_2) = (Z_1 \cup Z_2, \vec{n}_1+\vec{n}_2)$. We declare two colored sets $\bm{Z}_1, \bm{Z}_2$ to be disjoint iff $Z_1, Z_2$ are disjoint. This means in particular, that, in the definition of the norm above, the condition  $\sum_{\bm{Z} \ni x}$ can be   recast as $\sum_{(Z,\vec{n}), (Z,\vec{n}) \sim (\{x\},\vec{n}')}$, {for any $\vec{n}'$,  with the binary relation $\sim$ indicating that the colored sets are not disjoint. Such a formulation is necessary when we apply abstract cluster expansion results in {Lemma 1}.}

A potential $\Phi$ defines an many-body operator $H_\Phi$ that depends periodically on a variable $\vec\theta \in [0,2\pi)^m$, by
$$
H_\Phi(\vec\theta)=  \sum_{Z,\vec{n}} \Phi_{Z,\vec{n}} e^{i \vec{n} \cdot  \vec{\theta}}
$$
Since one can make the relation between potentials and many-body operators one-to-one (allowing for the addition of a constant to the many-body operator), we will in practice simply conflate $H$ and $H_\Phi$, and so we view the above norms $||\cdot ||_\kappa$ as a local norm on many-body operators.

\subsection{Proof of Theorem 1.}
We now give the proof of Theorem 1 in the main text. We take $m$, the number of components of $\vec\theta$, to be $m=1$ and so we write simply $\theta$ instead of $\vec\theta$. \\
{\bf Theorem 1. Approximate $U(1)$-conservation in Floquet systems.} \\
{\it Let $N$ (i) be a sum of local terms that mutually commute \& (ii) has integer spectrum, and $H(\theta)$ be a many-body Hamiltonian where $\theta$\,$\in$\,$[0,2\pi)$, with local norm $\|H(\theta)\|_{\kappa_0}$\,$<$\,$\infty$ for some $\kappa_0$\,$>$\,$0$.
 Let $\omega$\,$>$\,$0$ and define the local energy scale $\nu_0$\,$:=$\,$\max\{2\|H(\theta)\|_{\kappa_0},\omega\}$. We consider the Hamiltonian
 \begin{align}
     G(\theta) = \nu N + H(\theta)
     \label{eqn:G1}
 \end{align}
 (and correspondingly, the time-periodic (Floquet) Hamiltonian $G(t)$\,$:=$\,$G(\theta_t)$ with fundamental frequency $\omega$), where the amplitude $\nu$ is assumed large, specifically $\nu$\,$>$\,$C\nu_0$ for some constant $C$ depending only on $\kappa_0$  but   not the volume of the system.
 Explicitly it is given as
 \begin{align}
     C^{-1} = \min\left\{1, \frac{\kappa_0}{12\pi}, \frac{1}{2A}, \frac{x}{64\sqrt{2}}\kappa_0^2 \right\}
 \end{align}
 where \begin{align}
     A & = \left( \frac{216\pi}{\kappa_0^2}   +  \left( 1 + \frac{72\pi}{\kappa_0^2}   \right)\frac{4 \pi}{e\kappa_0}      \right), \nonumber \\  
     x & = \min\left\{\frac{1}{6\pi \kappa_0}, \frac{-(108\pi+4\pi\kappa_0/e) + \sqrt{(108\pi + 4\pi\kappa_0/e)^2 + 288 \pi \kappa_0/e} }{288 \pi \kappa_0/e} \right\}.
 \end{align}
 Then, there is a small unitary $e^{A(\theta)}$ effected by a quasilocal, antihermitian operator $A(\theta)$, such that the   unitary propagator corresponding to $G(t)$   can be written
 \begin{align*}
     U(t) = e^{A(\theta_t)} \mathcal{T}\exp\left(-i \!\! \int_0^t \!\!\! ds \nu N + D(\theta_s) + V(\theta_s) \right)e^{-A(\theta_0)},
 \end{align*}
 where $\mathcal{T}$ represents time-ordering, and $D(\theta)$\,$,$\,$V(\theta)$ are quasilocal, many-body Hamiltonians satisfying
 \begin{align}
     & \|D(\theta) - \langle H (\theta) \rangle \|_{\kappa} \leq C'(\nu_0/\nu), \label{eqn:estimate1} \\
     & \| V(\theta)\|_{\kappa} \leq \nu_0 2^{-n_*}, \\
     & [D(\theta),N] = 0.
 \end{align}
 Here $\kappa$\,$=$\,$\kappa_0/4$, $\langle\cdot\rangle$ represents the symmetrization operation $\langle O(\theta)\rangle$\,$=$\,$\frac{1}{2\pi} \int_0^{2\pi} d\phi e^{i \phi N} O(\theta) e^{-i \phi N} $, and \begin{align}
     n_* = \left\lfloor \frac{3  x \kappa_0^2}{32 \sqrt{2}} \left( \frac{\nu}{\nu_0} \right) \right\rfloor.
 \end{align}  $C'$  is a   numerical constant independent of volume.
 }
 \\

{\it Proof.} We relabel the initial $G(\theta),H(\theta) \mapsto G_0(\theta), H_0(\theta)$ and we will define renormalized operators $G_n(\theta),H_n(\theta)$. 
At each step we will also split the operator $H_n(\theta)$  into a term diagonal in $N$ and a term completely off-diagonal in $N$:
\begin{align}
    H_n(\theta) = D_n(\theta) + V_n(\theta)
\end{align}
where $D_n(\theta) := \langle H_n(\theta) \rangle$, $V_n(\theta):= H_n(\theta) - \langle H_n(\theta)\rangle$. Clearly, $[D_n(\theta),N] = 0$.

To define the renormalized Hamiltonians, at the $n$-th order, we introduce an antihermitian operator $A_n(\theta)$ defined via
\begin{align}
& A_n(\theta) = \frac{i \nu }{2\pi} \int_0^{\frac{2\pi}{\nu}} dt \int_0^t ds e^{is \nu N} V_n(\theta) e^{-i s \nu N}
\label{eqn:defining_A}
\end{align}
which satisfies
\begin{align}
&  [\nu N, A_n(\theta)] = -V_n(\theta).
\end{align}
We use this to define the next $G_{n+1}(\theta)$:
\begin{align}
G_{n+1}(\theta) & := e^{-A_n(\theta)} G_n(\theta) e^{A_n(\theta)} - i \omega e^{-A_n(\theta)} \partial_\theta e^{A_n(\theta)}   \\
& \equiv \nu N + H_{n+1}(\theta)  \\
& = \nu N + D_{n+1}(\theta) + V_{n+1}(\theta).
\end{align}
By introducing notation
\begin{align}
\gamma_n(O) &:= e^{-A_n} O e^{A_n} \\
\alpha_n(O) & := \int_0^1 ds e^{-s A_n} O e^{s A_n},
\end{align}
we can write $H_{n+1}(\theta)$ as 
\begin{align}
H_{n+1}(\theta) & = \left( \gamma_n(H_n(\theta)) + \gamma(\nu  N) - \nu N  \right) - i \omega \alpha_n(\partial_\theta A_n(\theta)) \nonumber \\
& = \gamma_n(H_n(\theta)) - \alpha_n([A_n(\theta),\nu N]) - i \omega \alpha_n(\partial_\theta A_n(\theta)) \nonumber \\
& = \gamma_n(D_n(\theta)) + (\gamma_n(V_n(\theta)) - V_n(\theta)) + (V_n(\theta) - [ A_n(\theta),\nu  N]  ) \nonumber \\
& ~~~  - ( \alpha_n([A_n(\theta),\nu N]) - [A_n(\theta),\nu N]) - i \omega \alpha_n(\partial_\theta A_n(\theta)) \nonumber \\
& = \gamma_n(D_n(\theta)) + (\gamma_n(V_n(\theta)) - V_n(\theta)) + ( \alpha_n(V_n(\theta)) - V_n(\theta)) - i \omega \alpha_n(\partial_\theta A_n(\theta)) 
\end{align}
It is useful to introduce 
\begin{align}
W_n(\theta) & := (\gamma_n(D_n(\theta)) - D_n(\theta)) + (\gamma_n(V_n(\theta)) - V_n(\theta)) + ( \alpha_n(V_n(\theta)) - V_n(\theta)) - i \omega \alpha_n(\partial_\theta A_n(\theta)) 
\end{align}
so that
\begin{align}
& D_{n+1}(\theta) = D_n(\theta) + \langle W_n(\theta) \rangle, \\
& V_{n+1}(\theta) = W_n(\theta) - \langle W_n(\theta) \rangle.
\end{align}
This concludes the recursion formulae defining the renormalization procedure of the Hamiltonians. The aim next is to provide bounds on the renormalized Hamiltonians. Note that formally, all manipulations have been similar to \cite{Abanin2017}, however, the main difference is that we have an additional term, the gauge potential $-i \omega e^{-A_n(\theta)} \partial_\theta e^{A_n(\theta)}$, which we have to account for. We assume  $A_n(\theta)$ is given by the choice Eq.~\eqref{eqn:defining_A}.
\\

We shall have to make use of two lemmas: \\
\noindent {\bf Lemma 1.} Let $Z(\theta),Q(\theta)$ be colored potentials on $S^1$ and assume that $3\|Q(\theta)\|_\kappa \leq \kappa - \kappa'$, with $0 < \kappa' < \kappa$. Then
\begin{align}
& \| e^{Q(\theta)} Z(\theta) e^{-Q(\theta)} - Z(\theta) \|_{\kappa'}  \leq \frac{18}{(\kappa -\kappa')\kappa'}\| Q(\theta)\|_\kappa \| Z(\theta) \|_\kappa, \\
& \| e^{Q(\theta)} Z(\theta) e^{-Q(\theta)} \|_{\kappa'} \leq \left( 1 + \frac{18}{(\kappa-\kappa') \kappa'} \|Q(\theta)\|_\kappa \right) \|Z(\theta)\|_\kappa.
\end{align}
 
\begin{proof}
Equivalent to Sec.~5.1 of \cite{Abanin2017}, replacing sets by colored sets. It is important to use the new notions of union $\cup$ and disjointness for colored sets. In particular, the notion of disjointness is crucial to set up the cluster expansion.   \end{proof}
 
\noindent{\bf Lemma 2.}  For $0 < \kappa' < \kappa$, 
\begin{align}
\| \partial_\theta O(\theta) \|_{\kappa'} \leq \frac{1}{e (\kappa-\kappa')} \| O(\theta) \|_\kappa.
\end{align}
 
\begin{proof}
Using $e^y>ye$ for any $y>0$, 
we have
\begin{align}
\| \partial_\theta O(\theta) \|_{\kappa'} & = \sup_x \sum_{Z\ni x, n} e^{\kappa'(|Z|+|n|) } \| i n O_{Z,n} \| \nonumber \\
& \leq \sup_x \sum_{Z \ni x,n} e^{\kappa'(|Z|+|n|)} \frac{1}{e(\kappa - \kappa')} e^{(\kappa - \kappa') |n|} \| O_{Z,n} \| \nonumber \\
& \leq \frac{1}{e (\kappa - \kappa') } \sup_x \sum_{Z \ni x, n} e^{\kappa(|Z|+|n|)} \| O_{Z,n} \| \nonumber \\
& \equiv \frac{1}{e(\kappa- \kappa') } \| O(\theta) \|_{\kappa}.    
\end{align}
\end{proof}

Armed with these lemmas, we now bound the renormalized Hamiltonians. Suppose there is a sequence of strictly decreasing decay constants $\kappa_0 > \kappa_1 > \kappa_2 > \cdots  > 0$, we then have
\begin{align}
& \| D_{n+1}(\theta) \|_{\kappa_{n+1}}  \leq \| D_n(\theta) \|_{\kappa_n} + w_n/2, \nonumber \\
& \| V_{n+1}(\theta) \|_{\kappa_{n+1}} \leq w_{n}, \nonumber \\
& \| D_{n+1}(\theta) -D_n(\theta) \|_{\kappa_{n+1}} \leq w_n/2, 
\end{align}
where $w_n = 2 \| W_n(\theta) \|_{\kappa_{n+1}}$.  Furthermore, 
\begin{align}
\|A_n(\theta) \|_\kappa   \leq \frac{\pi}{\nu} \| V_n(\theta) \|_\kappa
\end{align}
for any $\kappa > 0$. 
which follows from Eq.~\eqref
{eqn:defining_A} viewed as the pointwise (in $\bf Z$) definition of a colored potential.

Now from Lemma 1, provided we have 
\begin{align}
3 \| A_n(\theta) \|_{\kappa'_n} < \kappa'_n - \kappa_{n+1}
\end{align}
for some intermediate $\kappa_n'$ (to be determined) such that $\kappa_{n+1} < \kappa_n' < \kappa_n$, which we note is satisfied if
 $(   \frac{3\pi}{\nu} \| V_n(\theta) \|_{\kappa'_n} < \kappa_n' - \kappa_{n+1}
)
$
, then
\begin{align}
\| W_n(\theta) \|_{\kappa_{n+1}} & \leq \frac{18}{(\kappa_{n}' - \kappa_{n+1}) \kappa_{n+1}} \|A_n(\theta) \|_{\kappa_n'}  ( \|D_n(\theta)\|_{\kappa_n'} + 2 \| V_n(\theta) \|_{\kappa_n'}  )  + \| \omega \alpha_n(\partial_\theta A_n(\theta)) \|_{\kappa_{n+1}}  \nonumber \\
& \leq  \frac{18 \pi }{ \nu (\kappa_{n}' - \kappa_{n+1}) \kappa_{n+1}} \|V_n(\theta) \|_{\kappa_n}  ( \|D_n(\theta)\|_{\kappa_n} + 2 \| V_n(\theta) \|_{\kappa_n}  )  + \| \omega \alpha_n(\partial_\theta A_n(\theta)) \|_{\kappa_{n+1}} \nonumber \\
& \leq \frac{18 \pi }{ \nu (\kappa_{n}' - \kappa_{n+1}) \kappa_{n+1}} \|V_n(\theta) \|_{\kappa_n}  ( \|D_n(\theta)\|_{\kappa_n} + 2 \| V_n(\theta) \|_{\kappa_n}  ) \nonumber \\
& ~~~ + \omega \left(1 + \frac{18}{(\kappa'_n - \kappa_{n+1}) \kappa_{n+1}  }\|A_n(\theta) \|_{\kappa_n'} \right)\|\partial_\theta A_n(\theta) \|_{\kappa_n'} \nonumber \\ 
& \leq \frac{18 \pi }{ \nu (\kappa_{n}' - \kappa_{n+1}) \kappa_{n+1}} \|V_n(\theta) \|_{\kappa_n}  ( \|D_n(\theta)\|_{\kappa_n} + 2 \| V_n(\theta) \|_{\kappa_n}  ) \nonumber \\
& ~~~ + \omega \left(1 + \frac{18 \pi }{\nu (\kappa'_n - \kappa_{n+1}) \kappa_{n+1}  }\|V_n(\theta) \|_{\kappa_n} \right)\|\partial_\theta A_n(\theta) \|_{\kappa_n'}.
\end{align}
We now need to estimate $\|  \partial_\theta A_n(\theta) \|_{\kappa_{n}'}$. We first work out $\partial_\theta A_n(\theta)$:
\begin{align}
 \partial_\theta A_n(\theta) & =
\frac{i \nu }{2\pi} \int_0^{\frac{2\pi}{\nu }} dt \int_0^t ds e^{is \nu  N} \partial_\theta V_n(\theta) e^{-i s \nu  N}. 
\end{align}
 Therefore we have 
 \begin{align}
     \| \partial_\theta A_n(\theta) \|_{\kappa'_n} \leq \frac{\pi}{\nu}\| \partial_\theta V_n(\theta)\|_{\kappa'_n} \leq \frac{\pi}{\nu}\frac{1}{e(\kappa_n-\kappa'_n)} \| V_n(\theta)\|_{\kappa_n}
 \end{align}
 from Lemma 2. Plugging this in we have
\begin{align}
    \|W_n(\theta)\|_{\kappa_{n+1}} & \leq  \frac{18 \pi }{ \nu (\kappa_{n}' - \kappa_{n+1}) \kappa_{n+1}} \|V_n(\theta) \|_{\kappa_n}  ( \|D_n(\theta)\|_{\kappa_n} + 2 \| V_n(\theta) \|_{\kappa_n}  ) \nonumber \\
& ~~~ + \frac{\omega}{\nu} \left(1 + \frac{18 \pi }{\nu (\kappa'_n - \kappa_{n+1}) \kappa_{n+1}  }\|V_n(\theta) \|_{\kappa_n} \right)    \frac{\pi}{e(\kappa_n - \kappa_n')}   \|V_n(\theta) \|_{\kappa_n}. 
\end{align}
Now we make the choice that $\kappa'_n = (\kappa_n + \kappa_{n+1})/2$ so that 
\begin{align}
    \|W_n(\theta)\|_{\kappa_{n+1}} & \leq  \frac{36 \pi }{ \nu (\kappa_{n} - \kappa_{n+1}) \kappa_{n+1}} \|V_n(\theta) \|_{\kappa_n}  ( \|D_n(\theta)\|_{\kappa_n} + 2 \| V_n(\theta) \|_{\kappa_n}  ) \nonumber \\
& ~~~ + \frac{\omega}{\nu} \left(1 + \frac{36 \pi }{\nu (\kappa_n - \kappa_{n+1}) \kappa_{n+1}  }\|V_n(\theta) \|_{\kappa_n} \right)   \frac{2 \pi}{e(\kappa_n - \kappa_{n+1})}   \|V_n(\theta) \|_{\kappa_n}
\label{eqn:W}
\end{align}
and the requirement for Lemma 1 to hold can be satisfied if 
\begin{align}
    \frac{6\pi}{\nu} \| V_n(\theta) \|_{\kappa_n} < \kappa_n - \kappa_{n+1}.
\end{align}

These are the ultimate expressions and now our aim is to start an inductive process and choose the decay constants $\kappa_n$ appropriately. Let us first define $\kappa_1 = \kappa_0/2$. Then
\begin{align}
    \| W_0(\theta)\|_{\kappa_1} \leq \left( \frac{216\pi}{\kappa_0^2}   + \left( 1 + \frac{72\pi}{\kappa_0^2} \frac{\nu_0}{\nu} \right) \frac{4 \pi }{e\kappa_0}   \right) \frac{\nu_0^2}{\nu} \leq A \nu_0 \frac{\nu_0}{\nu}
\end{align}
(here we made use of the assumption $\nu > C \nu_0$).
%
The requirement (for Lemma 1 to hold) reads
\begin{align}
    \frac{6\pi \nu_0}{\nu} < \frac{\kappa_0}{2} \text{ or } \frac{\nu_0}{\nu} < \frac{\kappa_0}{12 \pi}
\end{align}
and is satisfied by similar assumption of the set-up of the problem.
Why did we take $\kappa_1 - \kappa_0$ to be independent of system parameters $\nu,\omega$? Well, in doing so, we have ensured $W_0(\theta)$ is small in $1/\nu$, so vanishes as $\nu \to \infty$ holding everything else fixed. 

Now we impose the inductive hypothesis that for some $n $,
\begin{align}
    & \| D_n(\theta) \|_{\kappa_n } \leq 2 \nu_0, \\
    &  \| V_n(\theta) \|_{\kappa_n} \leq \nu_0 \left(\frac{1}{2}\right)^n.
\end{align}
Clearly this is true for $n = 0$ by definition, and also true for $n = 1$, since the assumption $\nu_0/\nu < 1/(2A)$  guarantees $\|V_1(\theta)\|_{\kappa_1} \leq w_0 \leq A \nu_0 (\nu_0/\nu) \leq \nu_0 (1/2)$.

Plugging in the induction hypothesis into Eq.~\eqref{eqn:W} we have
\begin{align}
    \|W_n(\theta)\|_{\kappa_{n+1}} & \leq   \frac{\nu_0}{\nu} \left( \frac{108 \pi }{ (\kappa_{n} - \kappa_{n+1}) \kappa_{n+1}} +  \left(1 + \frac{36 \pi }{ (\kappa_n - \kappa_{n+1}) \kappa_{n+1}  } \frac{\nu_0}{\nu} \right) \left(  \frac{2\pi}{e(\kappa_n - \kappa_{n+1})} \right)  \right) \|V_n(\theta) \|_{\kappa_n}.
    %
\end{align}
Let us now impose the condition that
\begin{align}
    \frac{1}{(\kappa_n - \kappa_{n+1}) \kappa_{n+1}} \frac{\nu_0}{\nu} \leq x
    \label{eqn:x}
\end{align}
for an $x > 0$ to be determined.
Then
\begin{align}
    \| W_n(\theta) \|_{\kappa_{n+1}} \leq \left(108\pi x + \left(1 + 36\pi x \right)\left(  2 \pi \kappa_1 x/e \right) \right) \| V_n(\theta) \|_{\kappa_n} \leq \frac{1}{2} \|V_n(\theta)\|_{\kappa_n} \leq \nu_0 \left(\frac{1}{2} \right)^{n+1},
\end{align}
if we choose $x > 0$ to be at most the positive root of the quadratic equation
\begin{align}
     P(x',\nu_0/\nu) := & 108\pi x' + \left(1 + 36\pi x' \right)\left( 2\pi \kappa_1 x'/e \right)   -  \frac{1}{2} \nonumber \\
     = &(72   \pi \kappa_1/e)x'^2 + \left[108\pi + 2 \pi \kappa_1/e  \right]x' -  \frac{1}{2}.
     \label{eqn:P}
\end{align}
which yields
\begin{align}
 x = \frac{-(108\pi+4 \pi\kappa_0/e) + \sqrt{(108\pi+4\pi\kappa_0/e)^2 +288 \pi \kappa_0/e}}{288\pi \kappa_0/e}.
\end{align}

We also have to satisfy the requirement for Lemma 1, which reads $6\pi \nu_0 (1/2)^{n}/\nu < \kappa_n - \kappa_{n+1}$. The choice Eq.~\eqref{eqn:x} works, provided we take $x \leq \frac{1}{6\pi \kappa_0}$, since
\begin{align}
     \frac{6\pi }{\nu}\| V_n(\theta) \|_{\kappa_n} \leq 6\pi \frac{\nu_0}{\nu} \left( \frac{1}{2} \right)^n < 6\pi \frac{\nu_0}{\nu} < \frac{1}{\kappa_0 x} \frac{\nu_0}{\nu} < \frac{1}{\kappa_{n+1} x} \frac{\nu_0}{\nu} \leq (\kappa_n - \kappa_{n+1} ).
\end{align}
Therefore, our ultimate choice is
\begin{align}
    x = \min\left\{\frac{-(108\pi+4\pi\kappa_0/e) + \sqrt{(108\pi+4\pi\kappa_0/e)^2 +288 \pi \kappa_0/e}}{288\pi \kappa_0/e},\frac{1}{6\pi \kappa_0} \right\}.
\end{align}

We now define the decay rates on $n = 1,2,3,\cdots$ as
\begin{align}
    & \kappa_n:= \kappa(n) \text{ for } n = 1,2,3,\cdots \text{ where } \\
    & \kappa(y)^2 := \kappa_1^2 - 2 B \left( \frac{\nu_0}{\nu} \right)(y-1), \qquad y \in \mathbb{R}.
\end{align}
Now $\kappa(y)$ is a concave down function, so we have that
\begin{align}
    \kappa_n - \kappa_{n+1} \geq -\kappa'(n) =  \frac{B(\nu_0/\nu)}{\kappa_n}.
    \label{eqn:kappa1}
\end{align}
Moreover, assuming we only look at $n$s such that $\kappa_n \geq \kappa_1/2 = \kappa_0/4$ then
\begin{align}
    (\kappa_{n+1}/\kappa_n)^2 = 1 - \frac{2B(\nu_0/\nu)}{\kappa_n^2} \geq 1 - \frac{8B(\nu_0/\nu)}{\kappa_1^2}.
\end{align}
So if we impose 
\begin{align}
   \frac{8B(\nu_0/\nu)}{\kappa_1^2} \leq \frac{1}{2}
   \label{eqn:imposition}
\end{align}
we would then have
\begin{align}
    (\kappa_{n+1}/\kappa_n)^2 \geq \frac{1}{2}.
    \label{eqn:kappa2}
\end{align}
Combining Eq.~\eqref{eqn:kappa1} and \eqref{eqn:kappa2} we would have
\begin{align}
    \frac{1}{\kappa_{n+1} ( \kappa_n - \kappa_{n+1} ) } \leq \frac{\sqrt{2}}{B(\nu_0/\nu)}
\end{align}
so we should pick
\begin{align}
    B = \frac{\sqrt{2}}{x}.
\end{align}
Imposition Eq.~\eqref{eqn:imposition} therefore reads
\begin{align}
    \frac{\nu_0}{\nu} \leq \frac{x}{64\sqrt{2}} \kappa_0^2.
\end{align}
Therefore, the maximal $n_*$ to which the iteration procedure can be carried out to, is
\begin{align}
    n_* := \left\lfloor \frac{3x \kappa_0^2}{32\sqrt{2} } \left(\frac{\nu}{\nu_0} \right) \right\rfloor.
\end{align}
This gives the claimed bound on  $V(\theta)$ in the main text defined as $V(\theta) := V_{n_*}(\theta)$ (abusing a little, notation). Similarly, $D(\theta)$ defined as $D(\theta) := D_{n_*}(\theta)$  satisfies  $[D(\theta),N]=0$. Lastly the estimate Eq.~\eqref{eqn:estimate1}   follows from summing the bound on $\|D_{n+1}(\theta) - D_n(\theta) \|_{\kappa_{n}+1}$.  $\blacksquare$

\subsection{Proof of Theorem 2.}
We state Theorem 2 again. \\
{\bf Theorem 2. Approximate $\mathbb{Z}_n$-charge conservation in quasiperiodically-driven systems away from the high-frequency limit.}\\
{\it
Consider $N$ (i) a sum of local terms that  mutually commute and (ii) has integer eigenvalue spacings. Fix a non-zero integer $n$ and let $H(\vec\theta)$ be a many-body Hamiltonian on $\vec\theta$\,$\in$\,$[0$\,$,$\,$2\pi)^2$, assuming that $\|H(\vec\theta)\|_{\kappa_0}$ for some $\kappa_0$\,$>$\,$0$. 
We introduce a frequency vector $\vec\omega$\,$=$\,$(\nu$\,$,$\,$\omega)$, define $\nu_0$\,$:=$\,$\max\{2\|H(\vec\theta)\|_{\kappa_0},\omega\}$, and consider the Hamiltonian
\begin{align}
    G(\vec\theta) = \frac{\nu}{n} N + H(\vec\theta)
    \label{eqn:G2}
\end{align}
(and corresponding, the time-quasiperiodic Hamiltonian $G(t)$\,$:=$\,$G(\vec\theta_t)$). We take $\nu$\,$>$\,$C\nu_0$ for some constant $C$ depending on $\kappa_0$ but not on the system's volume.
Then, there is a small time-quasiperiodic unitary $e^{A(\vec\theta_t)}$ effected by a quasilocal, antihermitian operator $A(\vec\theta)$   such that the unitary propagator can be written  
\begin{align*}
    U(t) = e^{A(\vec\theta_t)} \mathcal{T} \exp\left(-i\!\! \int_0^t \!\!\! ds \frac{\nu}{n}N + D(\vec\theta_t) + V(\vec\theta_t) \right) e^{-A(\vec\theta_0)},
\end{align*}
where $D(\vec\theta), V(\vec\theta)$ are quasilocal Hamiltonians satisfying
\begin{align}
    & \| D(\vec\theta) - \langle H(\vec\theta)\rangle \|_{\kappa} \leq C'(\nu_0/\nu), \\
    & \| V(\vec\theta) \|_{\kappa} \leq \nu_0 2^{-n_*}, \\
    & [D(\vec\theta),g] = 0.
\end{align}
Here   $D(\vec\theta)$\,$=$\,$D'(\theta_2)$ has   dependence only on $\theta_2$\,$\in$\,$[0,2\pi)$, 
  $\kappa$\,$=$\,$\kappa_0/4$,  $g$\,$=$\,$e^{-i \frac{2\pi}{n}N }$ is a generator of the $\mathbb{Z}_n$ group satisfying $g^n$\,$=$\,$\mathbb{I}$,   $\langle \cdot \rangle$ is the symmetrization operation 
  $\langle O(\vec\theta)\rangle$\,$=$\,$\frac{1}{2\pi n}\int_0^{2\pi n} d\theta_1 e^{-i \frac{\theta_1}{n}N}O(\vec\theta)e^{i \frac{\theta_1}{n}N}$, 
  $n_*$\,$=$\,$\lfloor c(\nu/\nu_0)\rfloor$, and   $C',c$ are numerical constants.
}


{\it Proof.} To begin, we move into the rotating frame of $\nu N/n$ and base our analysis on the related Hamiltonian
\begin{align}
    H_0(\vec\theta)= e^{-i\frac{\theta_1}{n}N} H(\vec\theta) e^{i \frac{\theta_1}{n}N}.
\end{align}
In this formulation, the large amplitude  appears only as a driving frequency (of frequency $\nu/n$), but crucially $H_0(\vec\theta)$ now obeys the so-called twisted time-translation property \cite{PhysRevX.10.021032}:
\begin{align}
    H_0(\theta_1,\theta_2) = g H_0(\theta_1+2\pi,\theta_2) g^\dagger
\end{align}
where $g = e^{i \frac{2\pi}{n}N}$ is a $\mathbb{Z}_n$ generator, which satisfies $g^n = \mathbb{I}$.

For ease of convenience we rescale $\theta_1 \mapsto \theta_1/n$ so that $H_0(\vec\theta)$ is defined on the ``standard torus'' $\mathbb{T}^2 = [0, 2\pi)^2$, then the twisted time-translation property reads
\begin{align}
    H_0(\theta_1,\theta_2) = g H_0(\theta_1+2\pi/n,\theta_2) g^\dagger.
\end{align}

We define the operation
\begin{align}
    \langle O(\vec\theta) \rangle = \frac{1}{2\pi} \int_0^{2\pi} d\theta_1 O(\vec\theta) \equiv O'(\theta_2)
\end{align}
which we can interpret as performing a `Born-Oppenheimer' approximation (treating $\theta_1$ as the fast mode and integrating it out, while treating $\theta_2$ as slow and frozen).
We will introduce a sequence of small, frame transformations $e^{B_k(\vec\theta)}$, $k = 1,2,3,\cdots$ up to an optimal order $k_*$ to renormalize the Hamiltonians, getting $H_k(\vec\theta)$ at each stage, which we split   according to 
\begin{align}
    & D_k(\vec\theta) = \langle H_k(\vec\theta)\rangle \equiv D'_k(\theta_2) \\
    & V_k(\vec\theta) = H_k(\vec\theta) - \langle H_k(\vec\theta)\rangle.
\end{align}
Note $D'_k$ is only a function of $\theta_2$. Also, $[D'_k(\theta_2),g] = 0$ because, if $O(\theta_1,\theta_2)=g O(\theta_1+2\pi/n,\theta_2)g^\dagger $, then 
\begin{align}
g \langle O(\vec\theta)\rangle g^\dagger= \frac{1}{2\pi}\int_0^{2\pi} d\theta_1 gO(\theta_1,\theta_2)g^\dagger=
\frac{1}{2\pi}\int_0^{2\pi} d\theta_1 O(\theta_1-2\pi/n,\theta_2) =  \langle O(\vec\theta)\rangle.
\end{align}

The renormalization procedure at level $k+1$ involves defining a Hamiltonian $H_{k+1}(\vec\theta)$ from the the previous one at level $k$ via
\begin{align}
    H_{k+1}(\vec\theta) & = e^{-B_k(\vec\theta)} \left(H_k(\vec\theta) - i \frac{\nu}{n} \partial_{\theta_1} - i \omega \partial_{\theta_2} \right) e^{B_k(\vec\theta)} \nonumber \\
    & = e^{-B_k(\vec\theta)} \left(D_k(\vec\theta) + V_k(\vec\theta) - i \frac{\nu}{n} \partial_{\theta_1} - i \omega \partial_{\theta_2} \right) e^{B_k(\vec\theta)}.
\end{align}
We choose $B_k(\vec\theta)$ to satisfy
\begin{align}
    V_k(\vec\theta) - i \frac{\nu}{n} \partial_{\theta_1} B_k(\vec\theta) = 0,
\end{align}
which we take as solution
\begin{align}
    B_k(\vec\theta) = -i \frac{n}{2\pi \nu} \int_0^{2 \pi} d \phi \int_\phi^{\theta_1} d\theta'_1 V_k( \theta_1',\theta_2).
\end{align}
The reason for the outer integral is to ensure $A_k(\vec\theta)$ has also the twisted time-translation symmetry  (should $V_k(\vec\theta)$ have such a property too):
\begin{align}
    B_k(\theta_1 + 2\pi/n,\theta_2) & = -i \frac{n}{2\pi \nu} \int_0^{2 \pi} d \phi \int_\phi^{\theta_1 + 2\pi/n} d\theta'_1 V_k( \theta_1',\theta_2) \nonumber \\
    & = -i \frac{n}{2\pi \nu} \int_0^{2 \pi} d \phi \int_{\phi-2\pi/n}^{\theta_1} d\theta'_1 V_k( \theta_1' + 2\pi/n,\theta_2) \nonumber \\
    & = -i \frac{n}{2\pi \nu} \int_0^{2 \pi} d \phi \int_\phi^{\theta_1} d\theta'_1 g^\dagger V_k( \theta_1',\theta_2) g \nonumber \\
    & = g^\dagger B_k(\theta_1,\theta_2) g.
\end{align}
In such a case, $H_{k+1}(\vec\theta)$ is then guaranteed to also have a twisted time-translation symmmetry.

Now, introducing similar notation as used before
\begin{align}
\gamma_k(O) &:= e^{-B_k} O e^{B_k} \\
\alpha_k(O) & := \int_0^1 ds e^{-s B_k} O e^{s B_k},
\end{align}
we can write $H_{k+1}(\theta)$ as 
\begin{align}
H_{k+1}(\vec\theta)  = & \gamma_k(H_k(\vec\theta))  -i \frac{\nu}{n} \alpha_k(\partial_{\theta_1}B_k(\vec\theta)) - i \omega \alpha_k(\partial_{\theta_2} B_k(\vec\theta)) \nonumber \\ 
 = & \gamma_n(D_k(\vec\theta)) + \left(\gamma_k(V_k(\vec\theta) ) - V_k(\vec\theta) \right) 
+ \left( V_k(\vec\theta) -i \frac{\nu}{n}  \partial_{\theta_1}B_k(\vec\theta)  \right) \nonumber \\
&  
- \left( i \frac{\nu}{n} \alpha_k(\partial_{\theta_1}B_k(\vec\theta)) - i \frac{\nu}{n}  \partial_{\theta_1}B_k(\vec\theta)  \right)- i \omega \alpha_k(\partial_{\theta_2} B_k(\vec\theta)) \nonumber \\
& = \gamma_k(D_k(\vec\theta)) + (\gamma_k(V_k(\vec\theta)) - V_k(\vec\theta)) + ( \alpha_k(V_k(\vec\theta)) - V_k(\vec\theta)) - i \omega \alpha_k(\partial_{\theta_2} B_k(\vec\theta)) 
\end{align}
It is useful to also introduce similarly
\begin{align}
W_k(\vec\theta) & := (\gamma_k(D_k(\vec\theta)) - D_k(\vec\theta)) + (\gamma_k(V_k(\vec\theta)) - V_k(\vec\theta)) + ( \alpha_k(V_k(\vec\theta)) - V_k(\vec\theta)) - i \omega \alpha_k(\partial_{\theta_2} B_k(\vec\theta)) 
\end{align}
so that
\begin{align}
& D_{k+1}(\vec\theta) = D_k(\vec\theta) + \langle W_k(\vec\theta) \rangle, \\
& V_{k+1}(\vec\theta) = W_k(\vec\theta) - \langle W_k(\vec\theta) \rangle.
\end{align}
Then,
\begin{align}
    & \| B_k(\vec\theta)\|_{\kappa}  \leq \frac{n \pi}{\nu} \| V_k(\vec\theta)\|_{\kappa}, \\
    & \| \partial_{\theta_2} B_k(\vec\theta) \|_{\kappa'} \leq   \frac{n \pi}{\nu  }  \| \partial_{\theta_2} V_k(\vec\theta) \|_{\kappa'}
    \leq  \frac{n \pi}{\nu}\frac{1}{e(\kappa - \kappa')} \| V_k(\vec\theta) \|_{\kappa}
\end{align}
for any $0 < \kappa' < \kappa$. 
The mathematical setup is the same as that of Theorem 1, and therefore all bounds from the previous $U(1)$ case can be copied, verbatim, replacing $\nu \mapsto \nu/n$. In particular there is an optimal order $k_*$ going as $\nu/\nu_0$ to which the renormalization procedure can be carried out to, which minimizes $V_{k_*}(\vec\theta)$'s local norm.

We now state the   form of the unitary. There exists a frame transformation $e^{B(\vec\theta)} = e^{B_0(\vec\theta)}  e^{B_1(\vec\theta)}\cdots  e^{B_{k_*}(\vec\theta)}$ where $B_k(\vec\theta)$ are defined on $\mathbb{T}^2 = [0,2\pi)^2$ such that
\begin{align}
    U(t) = e^{-i \frac{\nu t }{n} N }e^{B(\vec\theta_t)} \mathcal{T} \exp\left(-i \int_0^t ds D_{k_*}(\vec\theta_s) + V_{k_*}(\vec\theta_s) \right) e^{-B(\vec\theta_0)}.
\end{align}
Here $D_{k_*}(\vec\theta) = D'_{k_*}(\theta_2)$ is only a function of $\theta_2$ and has a $\mathbb{Z}_n$ symmetry $[D_{k_*}(\vec\theta),g]= 0$.
Importantly, the flow here is of the form $\vec\theta_t = (\nu/n,\omega) t + \vec\theta_0$. 

If we want to work in the original  coordinates we simply scale back $\theta_1 \mapsto n \theta_1$. In the original coordinates $e^{B(\vec\theta)}$ is not invariant under translations by $2\pi$ in the $\theta_1$ direction, but a related object is:
\begin{align}
    e^{A(\vec\theta)} := e^{-i \frac{\theta_1}{n} N} e^{B(\vec\theta)} e^{i \frac{\theta_1}{n} N}.
\end{align}
To wit,
\begin{align}
    e^{A(\theta_1 + 2\pi,\theta_2)} & =  e^{-i \frac{\theta_1}{n} N} e^{-i \frac{2\pi}{n} N}  e^{B(\theta_1+2\pi,\theta_2)} e^{i \frac{2\pi}{n} N} e^{i \frac{\theta_1}{n} N} \nonumber \\
    & = e^{-i \frac{\theta_1}{n} N} g g^\dagger  e^{B(\vec\theta)} g^\dagger g e^{i \frac{\theta_1}{n} N} \nonumber \\
    & = e^{A(\vec\theta)}.
\end{align}
Therefore we have the final result
\begin{align}
    U(t) = e^{A(\vec\theta_t)}  \mathcal{T} \exp\left(-i \int_0^t ds  \frac{\nu}{n}N +  D_{k_*}(\vec\theta_s) + V_{k_*}(\vec\theta_s) \right) e^{-A(\vec\theta_0)}
\end{align}
where $\vec\theta_t = \vec\omega t + \vec\theta_0 = (\nu,\omega)t + \vec\theta_0$, the original flow (or the `original' driving frequencies).
Note we have `reinserted' $e^{-i \frac{\nu t}{n} N}$ into the time-ordered exponential at the expense of redefining $V_{k_*}(\vec\theta)$. 
However, in doing so, all objects are now defined on the standard torus $\mathbb{T}^2$ of the original problem. $\blacksquare$

In pedestrian terms, this says: (i) there is a small time-quasiperiodic change of frame $e^{A(\vec\theta_t)}$, such that dynamics is decomposed into two parts: (ii) there is a `large' overall oscillating envelope $e^{-i \frac{\nu t}{n} N}$, which realizes the generator $g$ of a $\mathbb{Z}_n$ group  raised to the power of $m$, at  time which is an $m$-mulitple of the period $T_1 = 2\pi/\nu$, and (iii) there is dynamics of a  time-periodic Hamiltonian $\frac{\nu}{n} N + D_{k_*}(\vec\theta_t) = \frac{\nu}{n} N + D'_{k_*}(\omega_2 t + (\theta_0)_2)$ with period $T_2 = 2\pi/\omega$.  Corrections to this arise due to exponentially small terms.

\subsection{Long-lived charge conservation}
Here we prove the statement that from Theorems 1 and 2 there is accompanying long-lived charge conservation in dynamics. To do so, it is convenient to make our setup a bit more explicit. We did not yet exclude that the local terms $N_S$ of $N$ grow unboundedly as the support set moves away from the origin of the lattice $\Lambda$. We do so now by requiring that $\sup_S \| N_S \| \leq n_0$ for some $n_0 \in \mathbb{N}$.  Additionally, we will use a simple lemma.\\[1mm]
\noindent{\bf Lemma 3.}  For any observable $O$ supported in $S$, and a Hamiltonian $H$, we have, for any $\kappa$
\begin{align}
\|[O,G] \| \leq  2 |S| \| O \|  \| H \|_\kappa 
\end{align}
 \begin{proof}
 By using the definition Eq.~\eqref{def: norm} directly, we find
$$\|[O,G] \| \leq  2 \| O \| \sum_{x\in S}  \sum_{(Z,n), Z \ni x}  \|H_{Z,n}\| \leq 2 \| H \|_\kappa 
 $$
 \end{proof}

Now for Theorem 1 we consider a time-dependent dressed charge
\begin{align}
    \tilde{N}(t) := e^{A(\theta_t)} N e^{-A(\theta_t)} 
    \label{eqn:dressing}
\end{align}
and  its Heisenberg time evolution $U(t)^\dagger \tilde{N}(t) U(t)$. This can be written as $e^{A(\theta_0)} U'(t)^\dagger N U'(t) e^{-A(\theta_0)}$ where $U'(t)$ is generated by the Hamiltonian 
\begin{align}
    H'(t) = D_{n_*}(\theta_t) + e^{-i \nu N t} V_{n_*}(\theta_t) e^{i \nu N t}.
\end{align}
The only symmetry-breaking processes are contained within the last term, so we can ask the question how similar Heisenberg time evolution of $N$ is under $H'(t)$, to Heisenberg time evolution just under $D_{n_*}(\theta_t)$.  
Formally, let the time evolution operator $U_1(t)$ be generated by $D_{n_*}(\theta_t)$, and $U_2(t)$ by $D_{n_*}(\theta_t) + e^{-i \nu N t} V_{n_*}(\theta_t) e^{i \nu N t}$. That is,
\begin{align}
    & i \partial_t U_1(t) = D_{n_*}(\theta_t) U_1(t), \\ 
    & i \partial_t U_2(t) = (D_{n_*}(\theta_t)+e^{-i \nu N t} V_{n_*}(\theta_t) e^{i \nu N t}) U_2(t),
\end{align}
with initial conditions $U_1(0) = U_2(0) = \mathbb{I}$. We can write the difference in Heisenberg time evolution as 
\begin{align}
    U_1(t)^\dagger N U_1(t) - U_2(t)^\dagger N U_2(t) & = N - U_2(t)^\dagger N U_2(t) \nonumber \\
    & = - \int_0^t ds U_2(s)^\dagger [U_1^\dagger(t-s) N U_1(t-s), e^{-i \nu N s }V_{n_*}(\theta_s)e^{i \nu N s}] U_2(s) \nonumber \\
    & = - \int_0^t ds U_2(s)^\dagger [N, e^{-i \nu N s }V_{n_*}(\theta_s)e^{i \nu N s}] U_2(s).
\end{align}
Bounding the expression using the standard operator norm, utilizing Lemma 3, gives
\begin{align}
     \| N - U_2(t)^\dagger N U_2(t) \| \leq 2 |\Lambda| t \nu_0 n_0 d^{R}  2^{-n_*}
 \end{align}
 where $R$ is the maximal range of $N$ and $d$ is the spatial dimension of $\Lambda$.
Therefore for $\tilde{N}$ we have
\begin{align}
    \frac{1}{|\Lambda|}\| \tilde{N}(0) - U(t)^\dagger \tilde{N}(t) U(t) \| \leq 2 t\nu_0 n_0 d^{R} 2^{-n_*}.
    \label{eqn:N_result}
\end{align}
This bound means that  $\tilde{N}(t)$ is conserved in Heisenberg time-evolution up to the prethermal timescale $\tau \sim e^{c' \nu/\nu_0}$. Focusing only on stroboscopic times $t = \mathbb{Z} T$ ($T$: period) gives the claimed result in the main text of long-lived conservation of $\tilde{N} := \tilde{N}(0)$ (in again a slight abuse of notation).
If we instead work with the original charge $N$, then we have an additioonal bounded error of $O(\nu_0/\nu)$ in Eq.~\eqref{eqn:N_result} arising from the small frame change in Eq.~\eqref{eqn:dressing}.

 For Theorem 2, it is less straighforward to express the charge conservation, because there is no infinitesimal generator associated to the symmetry group.  The most meaningful way to express that the charge is approximately conserved is by exhibiting that the dynamics of local observables is generated by a charge-conserving generator, up to an error that is very small in norm.  Such a statement can easily be deduced from the smallness of $V$: 
  Let $\widetilde U(t)$ be the unitary propagator $U(t)$ but with the error term $V$ set to zero. Then,by the Duhamel formula,  
 \begin{align}
    U(t)^\dagger OU(t)- \widetilde U(t)^\dagger O \widetilde U(t) = i \int_0^t ds   U(t)^\dagger U(s) [ V(\vec\theta_s) ,U(s)^\dagger O \widetilde U(s)]  U(s)^\dagger U(t).
\end{align}
 To bound the right-hand side, we invoke Lemma 3 and the Lieb-Robinson bound \cite{Lieb1972} to argue that $U(s)^\dagger O \widetilde U(s)$ is a local observable if $O$ is. 
 We refer to \cite{Abanin2017} for the details and we simply state the result, namely
 \begin{align}
   \|  U(t)^\dagger OU(t)- \widetilde U(t)^\dagger O \widetilde U(t) \| \leq   \nu_0 2^{-n_*}  (Ct+C')^{d} \|O\| |S| 
\end{align}
for some constants $C,C'$, and with $t$ the spatial dimension of the lattice $\Lambda$ and $S$ the support of $O$. 

In particular, if the  observable $O$ is symmetric, $[O,g]=0$, then $\widetilde U(t)^\dagger O \widetilde U(t) $ is $g$-symmetric and hence
\begin{align}
\| [U(t)^\dagger OU(t),g] \| \leq   \nu_0 2^{-n_*}  (Ct+C')^{d} \|O\| |S| ,
\end{align}
so that indeed, $O$ approximatively remains $g$-symmetric in the rotated frame.

\section{Extensions}

\subsection{Non-constant amplitude $\nu$}
Our theorems specify having the amplitude in front of the $N$ operator be constant: $\nu$ for Theorem 1 and $\nu/n$ for some fixed positive integer $n$ for Theorem 2. We can straightforwardly extend the domain of Theorems to encompass a large class of time-dependent amplitudes $\nu \mapsto \nu(t)$.
There are actually two ways to encompass such situations.

We focus   on the setting given in Theorem 1.
Case (i): Suppose 
\begin{align}
    \nu(t) = \bar{\nu}(1 + f(t))
    \label{eqn:nu_t}
\end{align} 
where $\bar{\nu}$ is a constant, and $f(t)$ is a sufficiently smooth,  time-periodic function with zero time-average satisfying $|f(t)| < 1$. We can always reparameterize time by defining \begin{align}
    t' = t + \int_0^t ds f(s).
\end{align}
Note that $t(t'+T)=t(t')+T$. 
Then, the solution of the Schr\"odinger equation 
\begin{align}
    i\partial_t U(t) = (\nu(t)N + H(t))U(t), \qquad U(0) = \mathbb{I}
\end{align}
can be obtained by solving the related equation
\begin{align} 
i \partial_{t'} U'(t') = \left( \bar{\nu} N + \left[ \frac{   H(t(t'))}{(1+f(t(t') ))} \right]  \right) U'(t'), \qquad U'(0) = \mathbb{I}
\end{align}
via $U(t) := U'(t'(t))$. Note that the Hamiltonian in the square parenthesis is time-periodic in $t'$-variables, so that we can apply Theorem 1 as quoted in the main text.

Case (ii): Suppose $\nu(t)$ is similarly as in Eq.~\eqref{eqn:nu_t} for some smooth $f(t)$. We can   simply eliminate $f(t)$ in Eq.~\eqref{eqn:nu_t} by going into the rotating frame associated with it,
\begin{align}
    U_0(t) = e^{-i \int_0^t ds f(s) N},
\end{align}
so that in the rotating frame, the Hamiltonian is 
\begin{align}
   G(t) =  \bar{\nu} N + U_0(t)^\dagger H(t) U_0(t).
\end{align}
Once again, Theorem 1 can be applied in such a formulation.

In similar fashion, we can apply Theorem 2 in situations where the amplitude $\nu \mapsto \nu(t)$ is time-dependent, specifically when it is time-periodic with period $T_1 = 2\pi/\nu$, as long as its time average is $\nu/n$.

\subsection{Extensions to multiple charge conservation,   drives with more fundamental frequencies}
We are not only limited to a single $U(1)$-charge conservation in Floquet systems nor a single $\mathbb{Z}_n$-charge conservation in a two-tone quasiperiodic system. Let us focus on extensions to the former scenario. We can in fact allow for {\it multiple} $U(1)$ charge conservation, by introducing $r > 1$ mutually commuting $N$ operators: $N_1, \cdots, N_r$, each appearing with amplitudes $\nu_1, \cdots, \nu_r$, so that the set-up is
\begin{align}
    G(\theta) = \sum_{k=1}^r \nu_k N_k + H(\theta).
\end{align}
(From this we   consider a Floquet system with Hamiltonian $G(t):=G(\theta_t)$, $\theta_t = \omega t + \theta_0$).
Explicitly, we need the condition that $\| H(\theta)\|_{\kappa_0}, \omega \ll \nu_k$ for all $k$, and we additionally require a Diophantine condition on the frequency vector $\vec\nu = (\nu_1, \cdots, \nu_r)$:
\begin{align}
    \frac{|\vec\nu\cdot \vec{n}|}{|\vec\nu|} \geq \frac{c}{|\vec{n}|^\gamma}
\end{align}
where $\vec{n} = (n_1,\cdots,n_r) \in \mathbb{Z}^r$ is any non-zero integer vector, for some $\gamma > 0$. This is so that one can control potentially dangerous resonances due to the small energy differences that can arise from a linear combination of multiple fundamental frequencies (see \cite{PhysRevX.10.021032} for more discussion).
It turns that for almost all choices of $\vec\nu$, the exponent satisfies $\gamma > r - 1$, with the constant $c$ depending on the choice of ratios $\nu_i/\nu_j$ but important not the overall frequency scale $|\vec\nu|$, as explained in \cite{PhysRevX.10.021032}. 
We expect that the proof  of Theorem 1 can be carried over with appropriate changes accounting for the multiple frequencies, adopting the techniques of \cite{PhysRevX.10.021032}, resulting in a prethermal lifetime of the multiple charges that is weaker than a pure exponential but still superpolynomially long in the large driving amplitude $|\vec\nu|$.

Furthermore, we can also extend the situation to allow for a drive with multiple fundamental frequencies (time-quasiperiodic driving), by promoting $\theta \mapsto \vec\theta = (\theta_1,\cdots,\theta_p)$ for some integer $p \geq 2$. That is, we can consider
\begin{align}
    G(\vec\theta) = \sum_{k=1}^r \nu_k N_k + H(\vec\theta)
\end{align}
with a flow $\vec\omega_t + \vec\theta_0$ defining a time-quasiperiodic Hamiltonian $G(t) := G(\vec\theta_t)$, where $\vec\omega = (\omega_1,\cdots,\omega_p)$ is a vector of drive frequencies. In this case, we require $\|H(\vec\theta)\|_{\kappa_0},\omega_l \ll \nu_k$ for all $l,k$, and a Diophantine condition on both frequency vectors $\vec\nu$ and $\vec\omega$, separately, to achieve multiple long-lived $U(1)$-charge conservation. 
Similar considerations also apply to the setting of Theorem 2 of the main text. We can consider a frequency vector $\vec\omega = (\nu_1,\cdots,\nu_r, \omega_1,\cdots,\omega_p)$ and a Hamiltonian
\begin{align}
    G(\vec\theta) = \sum_{k=1}^r \frac{\nu_k}{n_k} N_k + H(\vec\theta),
\end{align}
with some fixed set of non-zero integers $(n_1,\cdots,n_r)$, as well as a Diophantine condition on the set of frequencies $(\nu_1,\cdots,\nu_r)$ and $(\omega_1,\cdots, \omega_p)$, separately, to achieve a 
 result of multiple long-lived $\mathbb{Z}_n$ charge conservation in time-quasiperiodc drives with more than two fundamental frequencies. 

\bibliography{refs}